# Vector Spin Chirality Switching in Noncollinear Antiferromagnets

Aritra Dey[1,2], R. Bhuvaneswari[3,4], Sourav Chowdhury[5], Souvik Banerjee[1], Manisha Bansal[6], Smritiparna Ghosh[7], Anwesha Bera[1,2], Raktim Maity[6], Jayjit Kumar Dey[5], Weibin Li[8], Ashalatha Indiradevi Kamalasanan Pillai[9], Manuel Valvidares[8], Subhajit Roychowdhury[10], Magnus Garbrecht[9], Tuhin Maity[6], Umesh Waghmare[3] and Bivas Saha[1,2,11,*]

## Affiliations

[1]*Chemistry and Physics of Materials Unit, Jawaharlal Nehru Centre for Advanced Scientific Research, Bangalore 560064, India.*

[2]*International Centre for Materials Science, Jawaharlal Nehru Centre for Advanced Scientific Research, Bangalore 560064, India.*

[3]*Theoretical Sciences Unit, Jawaharlal Nehru Centre for Advanced Scientific Research, Bangalore 560064, India.*

[4]*School of Electrical and Electronics Engineering, SASTRA Deemed University, Thanjavur 613401, India.*

[5]*Deutsches Elektronen-Synchrotron (DESY), Notkestrasse 85, Hamburg 22607, Germany.*

[6]*School of Physics, Indian Institute of Science Education and Research Thiruvananthapuram, Thiruvananthapuram, Kerala 695551, India.*

[7]*UGC-DAE Consortium for Scientific Research, University Campus, Khandwa Road, Indore 452017, India.*

[8]*ALBA Synchrotron Light Source, E-08290 Cerdanyola del Vallès, Barcelona, Spain.*

[9]*Sydney Microscopy and Microanalysis, The University of Sydney, Camperdown, NSW 2006, Australia.*

[10]*Department of Chemistry, Indian Institute of Science Education and Research Bhopal, Bhopal, Madhya Pradesh 462066, India.*

[11]*School of Advanced Materials and Sheikh Saqr Laboratory, Jawaharlal Nehru Centre for Advanced Scientific Research, Bangalore 560064, India*

*Correspondence to: bsaha@jncasr.ac.in and bivas.mat@gmail.com

**Fully Referenced Summary Paragraph:**

Spin chirality provides a powerful route to control magnetic and topological phases in materials, enabling next-generation spintronic and quantum technologies[1–4]. Coplanar noncollinear antiferromagnets with Kagome lattice spin geometries host vector spin chirality (VSC)[5], the handedness of spin arrangement, and offer an excellent platform for chirality-driven phase control [6,7]. However, the microscopic mechanisms governing VSC switching and its coupling to magnetic order, electronic structure, and quantum geometry remain elusive, with experimental evidence still lacking. Here, we present conclusive experimental evidence of temperature-driven VSC switching in an archetypal noncollinear antiferromagnetic manganese chromium nitride ($Mn_3CrN$) epitaxial thin films. The VSC switching induces a concomitant quantum-geometric and Lifshitz transition, manifested through a pronounced peak in anomalous Hall conductivity remanence, a metal-insulator-like crossover in longitudinal resistivity, and a distinct evolution of x-ray magnetic circular dichroic signal. The reversal of VSC reconstructs the spin configuration, Fermi surface topology and Berry curvature, marking a unified magnetic-electronic-quantum geometric transition. This emergent behaviour, captured through magneto-transport and magneto-optic measurements, and supported by first-principles theory establish VSC as an active control knob for chirality-driven phase engineering and the design of multifunctional quantum devices [7–11].

**Main**

Spin chirality, a fundamental geometric property arising from the noncollinear and/or noncoplanar arrangement of localized spins, has emerged recently as an important concept in condensed matter physics [1–4]. Unlike the conventional magnetic order parameters that describe spin alignment, spin chirality encapsulates the handedness of spin configurations, giving rise to profound topological and transport phenomena. It couples the spin, charge, and orbital degrees of freedom, leading to unconventional effects such as the topological Hall effect, anomalous magnetotransport, and emergent electromagnetic fields [6,12]. Moreover, the interplay between spin chirality and spin–orbit coupling (SOC) enables the stabilization of complex spin textures, including skyrmions, hedgehogs, and vortices, that serve as nanoscale information carriers [13,14]. Therefore, controlling spin chirality through external stimuli such as temperature, strain, electric fields, or light offers powerful means to engineer new magnetic and topological phases and holds promise for next-generation spintronic, quantum, and neuromorphic devices [7–11].

Coplanar noncollinear antiferromagnets (N-AFMs) with Kagome-lattice spin structures have emerged as an ideal platform to explore the VSC. Unlike collinear antiferromagnets, N-AFMs combine ultrafast spin dynamics, non-volatility, and immunity to stray fields with efficient electrical readout like ferromagnets [14]. The breaking of time-reversal symmetry in coplanar N-AFMs further induces Berry curvature–driven anomalous responses, bridging antiferromagnetic and topological behaviour [4]. However, despite these compelling attributes, the microscopic mechanisms that govern VSC switching and bind it to magnetic, electronic, and quantum-geometric degrees of freedom remain unresolved [6,7], owing to the lack of direct experimental evidence.

In coplanar noncollinear triangular spin systems (see Fig. 1a), where $S_1$, $S_2$, and $S_3$ are three neighbouring coplanar spins, the VSC [5] is defined as-

$$\kappa = \frac{2}{3\sqrt{3}} \sum_{\langle ij \rangle} [S_i \times S_j]_z = \begin{cases} +1, & \text{(Right handed (direct) chirality)} \\ -1, & \text{(Left handed (inverse or staggered) chirality)} \end{cases} \quad (1)$$

where the sum runs over nearest-neighbor spin pairs in the triangle, the subscript $z$ denotes the out-of-plane component of the cross product, and $S_i = S\,\hat{n}_i$, with $\hat{n}_i$ as the unit vector specifying the spin direction at site $i$. Importantly, the unit magnitude of κ signifies a well-defined chiral order, while the sign distinguishes between the two chiral states, right-handed (direct) chirality with κ = +1 and left-handed (inverse or staggered) chirality with κ = -1. This chiral order parameter alters the electronic structure and Fermi surface topology leading to Lifshitz transition [15–17], modifies the quantum geometry of electronic states including Berry curvature [18] distribution (see Fig. 1b and 1c) and controls the emerging topological transport signatures such as the AHE [5,19,20]. Furthermore, recent theoretical studies have predicted that some N-AFMs can host topologically nontrivial electronic structures, including Weyl nodes and a nodal-ring-to-Weyl-point transition [21] driven by VSC.

N-AFMs based-on the cubic antiperovskite nitride $Mn_3XN$ [22,23] (X: transition metal or main group elements), such as $Mn_3GaN$, $Mn_3ZnN$, $Mn_3PtN$, and $Mn_3IrN$ [24], exhibit strong spin–orbit coupling (SOC), pronounced anomalous Hall conductivity and magneto-optic responses.

Notably, Mn$_3$PtN and Mn$_3$IrN possess high Néel temperatures ($T_N$) [24] and significant spin–lattice coupling, underscoring their potential for technological applications. In terms of their spin orientation, Mn$_3$XN-based N-AFMs support distinct spin configurations, most prominently characterized by the $\Gamma_{4g}$ and $\Gamma_{5g}$ magnetic orderings (see Supplementary Fig. S1c in Supplementary Materials (SM)). In the $\Gamma_{4g}$ phase, the magnetic structure/ordering is invariant under the $C_{3[111]}$ and combined action of time-reversal ($T$) and mirror ($M$) symmetries across the $M_{\bar{1}01}$, $M_{1\bar{1}0}$ and $M_{01\bar{1}}$ planes [25,26]. In contrast, the magnetic structure of $\Gamma_{5g}$ phase is invariant only under $M$ and $C_{3[111]}$ symmetries ($T$ is absent) [25]. Importantly, within each chiral phase (either κ = +1 or κ = -1), a continuous spin rotation by an angle $\theta$, while preserving the $120^0$ relative angles between spin pairs with $C_{3[111]}$ symmetry, enables a manifold of nearly degenerate magnetic configurations, further enriching the magnetic phase space. $\Gamma_{4g}$ phase permits an anomalous Hall response with SOC due to $T * M$ symmetries, while the presence of only mirror symmetries ($M_{\bar{1}01}$, $M_{1\bar{1}0}$ and $M_{01\bar{1}}$) in $\Gamma_{5g}$ phase leads to vanishing AHE [27]. Yet, despite their promise, the practical implementation of Mn$_3$XN-based N-AFMs is challenged by competing magnetic phases, growth complexities, particularly in thin-film form, and limited control over spin chirality.

In this work, we present conclusive experimental evidence of temperature-induced VSC switching in epitaxial, single-crystalline Mn$_3$CrN thin film, driving a distinct spin reorientation accompanied by a Lifshitz transition. The spin reconfiguration is further correlated with changes in the system's quantum geometry, leading to a modulation of the AHC remanence via Berry curvature redistribution and evolution of the x-ray magnetic circular dichroic (XMCD) signal. Moreover, the epitaxial growth of Mn$_3$CrN opens its prospects for integration into next-generation spintronic and quantum information devices.

**Spin-Chirality Switching and Magneto-Optic Signatures**

Epitaxial and near-stoichiometric ~ 70 nm thick Mn$_3$CrN thin films are deposited inside an ultra-high vacuum sputtering deposition chamber at a base pressure of $1 \times 10^{-9}$ Torr at 470°C substrate temperature. Mn$_3$CrN films adopt a cubic antiperovskite rocksalt crystal structure, with space group $Pm\bar{3}m$ and {111} set of Kagome planes. All the films grow with (002) orientations on (001) MgO and (001) SrTiO$_3$ substrates and are single-crystalline without any significant extended defects. A detailed discussion on the growth process and structural characterization is presented later in the manuscript as well as in the SM.

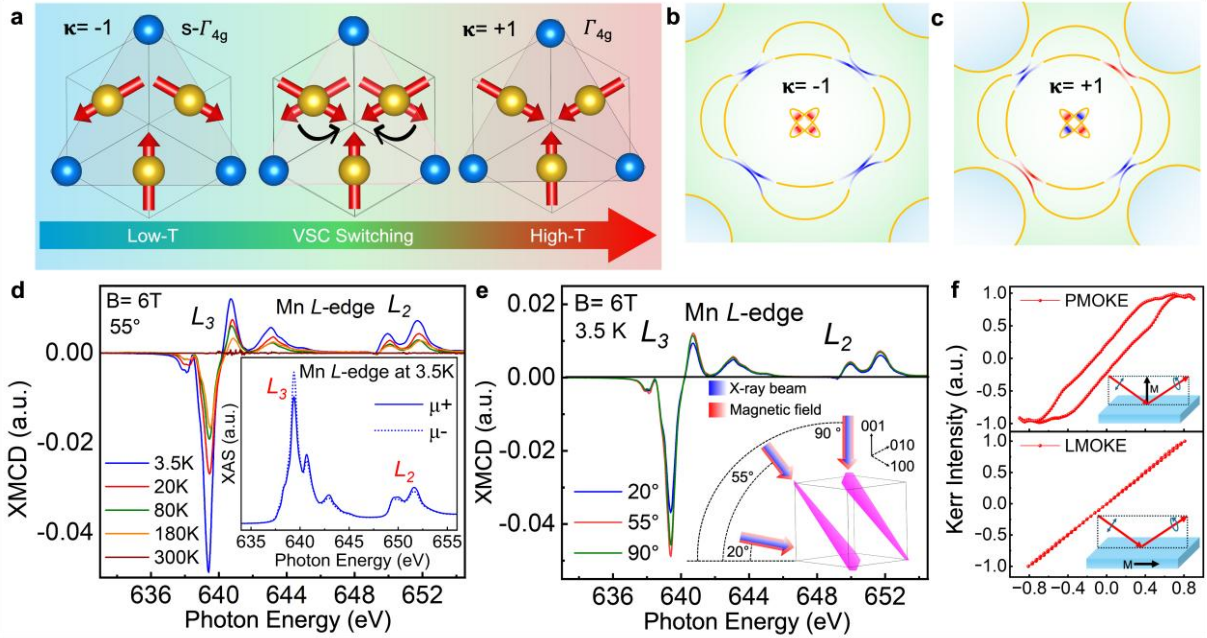

**Fig. 1| Vector Spin Chirality Switching in Noncollinear Antiferromagnets. a**, Schematic of the evolution of VSC, defined as κ, with temperature, starting from a negative chirality state (κ = –1) at low temperatures into a positive chirality state (κ = +1) at high temperatures. Blue spheres represent Cr atoms, while yellow spheres denote Mn atoms with spin directions indicated by red arrows. **b, c**, Illustration of the concomitant Fermi surface topology and Berry curvature evolution accompanying the VSC switching. The yellow contours depict the Fermi surfaces, while red/blue regions mark Berry curvature hot spots. In the κ = –1 state, the Fermi surface exhibits pronounced Berry curvature hot spots, whereas in the κ = +1 state the Fermi surface topology is reshaped, and the net Berry curvature vanishes due to symmetry constraints. **d**, Temperature-dependent XMCD at the Mn $L_{3,2}$-edges showing a maximum finite dichroic signal at 3.5 K from the κ = –1 chirality state. The XMCD signal progressively decreases with increasing temperature and vanishes at 300 K due to symmetry-enforced cancellation of XMCD signal in the κ = +1 chirality state. The inset shows the XAS signal at 3.5K, highlighting the $\mu^+ - \mu^-$ difference that gives rise to XMCD. **e,** Angle-dependent XMCD at ~ 20°, 55°, and 90° at 3.5 K, with consistent line-shape but varying intensity, showing the strongest signal at ~ 55° (along the Kagome plane). The inset shows the schematic of the angle-dependent XMCD measurement, where the Kagome planes are shown in pink and both the incident X-ray beam and magnetic field are applied parallel to each other. **f**, Polar (P) and longitudinal (L) magneto-optical Kerr effect (MOKE) measurements (upper and lower panel respectively), with strong response in the out-of-plane geometry (P-MOKE), indicating non-vanishing Berry curvature. Insets of PMOKE and LMOKE show the schematic of the measurement configuration in both geometries.

Temperature and angle-dependent XMCD measurement at the Mn $L_{3,2}$ edges (see Fig. 1d and 1e, respectively) provide direct conclusive evidence of temperature-driven VSC transition in Mn$_3$CrN. Temperature-dependent spectra, recorded at ~55° (along the (111) Kagome plane) under a 6 T field, show a maximum finite dichroic signal at 3.5 K, consistent with stabilization of the κ = -1 chirality state. At the same time, the response vanishes gradually at 300 K due to symmetry-enforced cancellation in the κ = +1 chirality state. This behaviour agrees with theoretical predictions that the XMCD signal vanishes upon the κ = –1 to κ = +1 chirality transition [28]. Angle-dependent measurements under a 6 T field and at 3.5 K further highlight the coupling between the incident helicity of the X-ray and the Kagome spin texture. The maximum XMCD intensity, occurs at ~ 55°, where the beam direction aligns with the Kagome spin plane. At the same time, reduced signals at 20° and 90° reflect only the respective projections.

Element-specific analysis confirms that Mn alone contributes to the XMCD, whereas Cr shows no magnetic response, establishing Mn as the sole source of magnetization. The Mn XAS line-shape matches that of Mn$^{2+}$ ions [29] (see the inset of Fig. 1d), and using the XMCD sum rules [30] (see XAS & XMCD measurements in Methods and Extended Data Fig. 1 for details), the spin moment ($s_Z$) and orbital moment ($l_Z$) are determined as -1.048 and 0.0248 μ$_B$/Mn-ion, respectively, yielding a total moment of -1.0232 μ$_B$/Mn-ion at 3.5 K. Complementary XAS at the N and O $K$-edges (see section E in SM) further verifies stoichiometry, with pronounced Mn–N hybridization at the N $K$-edge and negligible O $K$-edge intensity, ruling out oxygen-related secondary phases. Thus, the angle dependent anisotropic XMCD response, maximized along the Kagome plane and suppressed away from it, together with its chirality-dependent temperature evolution, provides spectroscopic evidence for a robust noncollinear antiferromagnetic ground state and VSC transition.

Polar MOKE (PMOKE) [31–33] (see Fig. 1f, top panel), measured with the magnetic field applied along the out-of-plane [001] direction, exhibits a pronounced hysteresis loop with clear saturation, reflecting the Berry-curvature–driven magneto-optical response[27,34,35]. This behaviour is consistent with the AHE (as shown in Fig. 2b), where a finite hysteresis arises from the underlying Berry curvature distribution. In contrast, longitudinal MOKE (LMOKE)[31–33] (see Fig. 1f, bottom panel), measured with the magnetic field applied along the in-plane [100] direction, shows negligible loop opening and a nearly linear field dependence. This response is analogous to the planar Hall effect [36], which occurs when both the current and magnetic field lie in the sample plane.

**Chirality-Driven Electronic and Magneto-Transport Transitions**

Temperature-dependent longitudinal resistivity ($\rho_{xx}$) increases monotonically from 30K to ~ 150K, beyond which a pronounced downturn is observed at higher temperatures (see Fig. 2a). Such changes in the positive-to-negative temperature coefficient of resistivity indicates a metal to insulator-like transition [37], traditionally observed in transition metal oxides such as $VO_2$ and other strongly correlated electronic systems [38,39]. However, unlike any previous precedence, this observed electronic transition in $Mn_3CrN$ occurs due to a temperature-driven reconfiguration of the magnetic ground state, wherein the VSC undergoes a transition from a left-handed ($\kappa = -1$) to a right-handed ($\kappa = +1$) chirality configuration (see Fig. 1a) [21] and concomitant change in the electronic band structure and Fermi surface topology, underscoring a Lifshitz Transition. These observed transport signatures, in conjunction with magneto-optic measurements as well as first-principles modelling presented subsequently, support the VSC switching that modulates the electronic structure. Note that the transition is unambiguous, with a distinct yet modest change in $\rho_{xx}$ across the transition temperature. At lower temperatures (< 35K), $\rho_{xx}$ increases slightly due to the disorder-induced electronic localization [40] (see section F in SM).

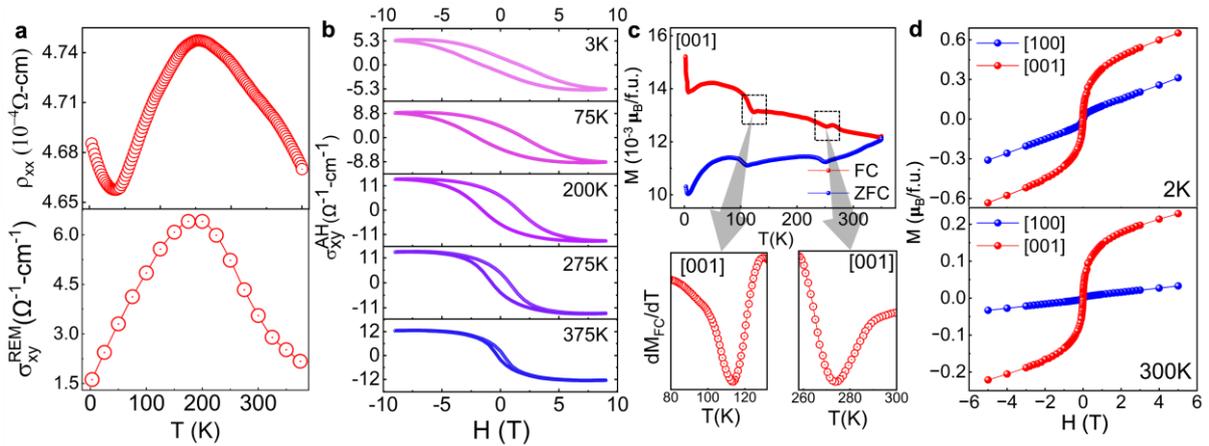

**Fig. 2| Electronic, magnetotransport and magnetic measurements of Mn₃CrN. a,** Longitudinal resistivity ($\rho_{xx}$) as a function of temperature showing (upper panel) a metallic to insulating-like transition at ~ 150 K characterized by the positive and negative temperature coefficient of resistivity, respectively, due to the temperature-induced VSC transition. Below 50K, $\rho_{xx}$ increases due to electron localization. Remnence

in anomalous Hall conductivity ($\sigma_{xy}^{REM}$) exhibiting (lower panel) a maximum at ~ 150K due to a drastic change in Mn$_3$CrN's Berry curvature. **b**, Anomalous Hall conductivity ($\sigma_{xy}$) as a function of the magnetic field (H), demonstrating a consistent anomalous Hall effect across a wide temperature range from 3 K to 375 K. Postive and negative $\sigma_{xy}^{REM}$ values at H=0 originates due to the Berry curvature polarity switching. **c**, Field cooled (FC) and zero-field cooled (ZFC) magnetization curves measured at 100 Oe showing distinct spin reorientation transitions as a function of temperature. Lower panel with a derivative (dM$_{FC}$/dT) clearly marks the commencement of spin transition. **d**, Field-dependent magnetization (M–H) measured along the in-plane (IP) and out-of-plane (OOP) directions at 2 K (top) and 300 K (bottom), showing antiferromagnetic nature of the film with a clear magnetic anisotropy between IP and OOP orientations, consistent with the easy-plane spin structure of the film.

The role of VSC switching in inducing $\rho_{xx}$ metal to insulator-like transition is further corroborated by the temperature-dependent AHC remanence ($\sigma_{xy}^{REM}$) determined via the relation $\sigma_{xy} = \frac{\rho_{yx}}{(\rho_{xx}^2 + \rho_{yx}^2)}$ [41,42], where, $\rho_{xx}$ and $\rho_{yx}$ represent the longitudinal and transverse resistivities, respectively. $\sigma_{xy}^{REM}$ exhibits a pronounced peak ~150K (see Fig. 2a) signaling at changes in the quantum geometry and coincides with the $\rho_{xx}$ electronic transition due to VSC switching. Above this transition temperature, $\sigma_{xy}^{REM}$ decreases gradually, consistent with a reduction in the net Berry curvature as well as enhanced electron-electron and electron-phonon scattering. Notably, the peak in $\sigma_{xy}^{REM}$ is not accompanied by any structural changes or distortions in Mn$_3$CrN detectable via synchrotron-radiation temperature-dependent high-resolution x-ray diffraction (HRXRD) (see section A in SM) indicating a temperature-driven transition between distinct noncollinear antiferromagnetic states characterized by different VSC.

To gain further insight into the VSC switching, magnetic-field-dependent AHC is measured across 2K-375K temperature range. Fig. 2b shows a robust and persistent AHE, with $\sigma_{xy}^{REM}$ remaining finite and sign-consistent across the entire temperature range. This unbroken AHE signal strongly suggests the absence of a transition purely into a mirror-symmetric magnetic configuration, such as the $\Gamma_{5g}$ state, which would otherwise lead to vanishing Berry curvature contributions and suppression of the AHE [25]. Instead, the sustained AHE is indicative of a transition between κ = -1 and κ = +1 chiral configurations, specifically from a left-handed staggered spin structure at low temperatures to a right-handed configuration at higher temperatures, confirmed with energetics from first principle DFT analysis. Further, a linear dependence of $\rho_{yx}^{AH}$ on $\rho_{xx}^2$ in the low-temperature region with intrinsic anomalous Hall conductivity of ≈ 41 S cm$^{-1}$ indicates that the AHE originates from the Berry curvature of the electronic structure (see section G in SM) [42,43]. Comparable AHE response along with the peak in $\sigma_{xy}^{REM}$ is also observed in Mn$_3$CrN films deposited on (001) SrTiO$_3$ substrates, further confirming the substrate-independent nature of the underlying VSC switching and topological signatures (see section H and I in SM).

The magnetic anomalies observed at ~ 115K and 265K in the temperature-dependent field cooled (FC) and zero-field cooled (ZFC) magnetization measured at 100 Oe field (see Fig. 2c, upper panel) occur well-within the magnetically ordered antiferromagnetic phase of Mn$_3$CrN, with $T_N$ > 300K. The lower-temperature transition around 115K corresponds to the commencement of a spin reorientation or switching of the VSC, which leads to the complete chirality switching transition observed near 265K (see Fig. 2c, upper panel). This staggered progression highlights the multi-stage evolution of the magnetic order, where a chirality-preserving magnetic state below 115K transitions into another chirality-active regime across 265K, elucidating its first-order phase transition nature (see Fig. 2c). The concomitant evolution of $\sigma_{xy}^{REM}$, $\rho_{xx}$ and magnetic measurements thus provides compelling evidence for a concomitant magnetic–electronic–quantum geometric transition governed by the reconfiguration of spin chirality.

The field-dependent magnetization (M–H) measured along the out-of-plane ([001]) and in-plane ([100]) directions exhibits negligible coercivity, nonlinearity, non-saturating behaviour at high fields, and the absence of hysteresis, consistent with the noncollinear antiferromagnetic nature of Mn$_3$CrN (see Fig. 2e and Extended Data Fig. 2). A pronounced anisotropy is observed between the two orientations, where the out-of-plane configuration shows a consistently larger magnetization compared to the in-plane direction. This anisotropy [44,45] originates from the geometric projection of the applied magnetic field onto the (111) Kagome planes. Out-of-plane fields project with a larger effective component ($\propto \sin 54.7°$) onto the Kagome plane, thereby coupling more strongly to the noncollinear spin texture. In contrast, in-plane fields couple less effectively ($\propto \cos 54.7°$) due to their smaller overlap with the Kagome plane. These anisotropic saturation characteristics underscore the directional sensitivity of spin interactions in Mn$_3$CrN and reflect the intrinsic magnetic anisotropy associated with its chiral antiferromagnetic ground state.

**First-Principles Mechanism of Chirality Reversal**

First-principles density functional theory (DFT) calculations [46,47] on the noncollinear magnetic phases of Mn$_3$CrN, incorporating SOC (see Computational Details in Methods), are performed to uncover the microscopic origin of the experimentally observed concomitant magnetic-electronic-quantum geometric transition. To capture lattice parameters and magnetic moments of Mn atoms accurately, a Hubbard U correction of 2 eV was used for on-site correlations of Mn-*3d* orbitals, accounting for moderate electronic correlations (see Extended Data Fig. 3a). DFT-estimated total energy difference between the magnetic phases of Mn$_3$CrN reveals that the left-handed chiral state ($\kappa = -1$) constitutes the low-temperature ground state, with a significant energy difference compared to the right-handed counterpart ($\kappa = +1$) (see Extended Data Table 1). Notably, within each chiral phase, the total energy change is negligible as the system transitions between $\Gamma_{4g}$ and $\Gamma_{5g}$ magnetic configurations, indicating a degeneracy between these magnetic orderings in terms of energetics (see Extended Data Fig. 3 and section L and M in SM). Consistently, the electronic structure and density of states also remain unchanged during this spin rotation, further confirming their energies and electronic degeneracy. In contrast, the inter-chirality transition between $\kappa = -1$ (low-temperature) and $\kappa = +1$ (high-temperature) phases lead to marked differences in electronic structure (see Figs. 3a, b). A small moment is induced on the Cr atom in the $\kappa = -1$ state, which enhances the stability of the $\kappa = -1$ staggered ground-state configuration and produces a distinct change in electronic structure of states with distinct VSC and with notable splitting of bands at Γ. However, given its tiny magnitude, XMCD measurements could not conclusively capture such moments on Cr atoms. The microscopic origin of these phenomena is traced to a monopole-like source–sink spin divergence pattern (see section J in SM).

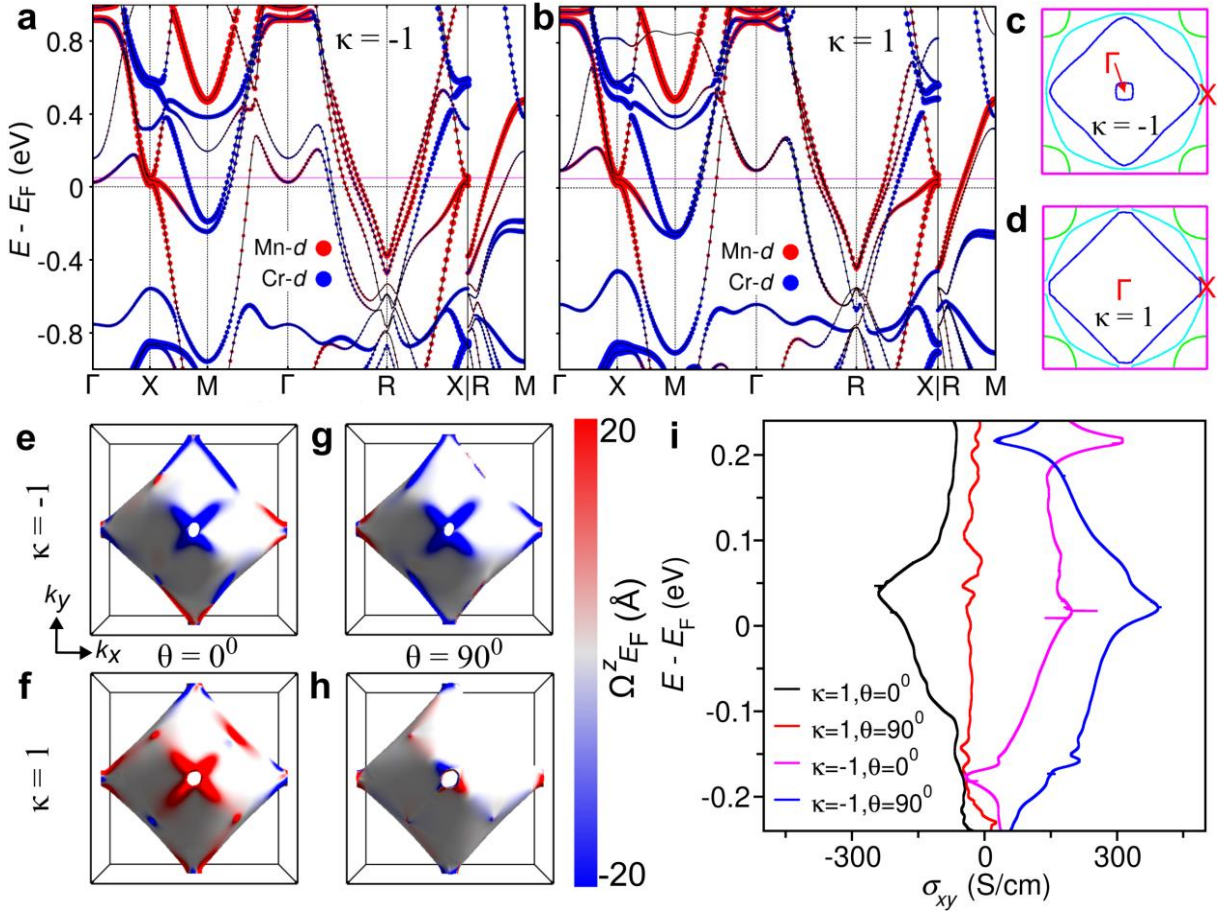

**Fig. 3| Chirality-dependent electronic structure and reconstruction of Fermi surface topology, and Berry curvature of Mn₃CrN. a**, **b**, Orbital-resolved band structures of the N-AFM states with (**a**) κ = −1 and (**b**) κ = +1, showing metallic character dominated by Mn-*d* and Cr-*d* orbitals. In the κ = −1 state, a weak induced magnetic moment appears on Cr atom (see Extended Data Fig. 3). Magenta lines indicate the energy slice used for Fermi surface analysis. **c**, **d**, Fermi surface cross-sections at 0.05 eV above $E_F$ for (**c**) κ = −1, revealing an electron pocket at Γ, which is missing in (**d**) κ = +1, consistent with a Lifshitz transition. **e–h**, Berry curvature $\Omega^z$ distributions on representative Fermi surface sheets for κ = −1 and κ = +1 under varying spin rotation angle θ, with θ = 0° and 90° corresponding to the $\Gamma_{4g}$ and $\Gamma_{5g}$ phases, respectively. **i**, Anomalous Hall conductivities $\sigma_{xy}$ obtained from Brillouin-zone integration of $\Omega^z$ for both chirality states.

Weak changes in the experimentally measured $\rho_{xx}$ as a function of temperature (see Fig. 1d), mimicking an electronic transition from κ = -1 to κ = 1 state can be correlated to the additional electron pocket accessible at energy slightly above $E_F$ (at 0.05 eV) in the κ = -1 state (see Fig. 3c). On contrary, no such pockets are accessible for κ = 1 state (see Fig. 3d). This chirality-dependent splitting and shifting of bands to lower accessible energies is primarily driven by subtle changes in Cr-*3d*-orbital hybridization. Moreover, our calculations show that the canting of the spins out of (111) plane (by angle $\theta_{out}$) is not energetically favorable, as shown in the SM section K.

From a topological standpoint, the chirality parameter κ acts as a thermodynamic and topological control knob that governs both the global symmetry and Berry curvature distribution. AHE exists in N-AFM coplanar systems in the presence of SOC that relaxes the combined effect of $T$ and spin-rotation ($R_S$) symmetries, provided the effect is allowed under other magnetic point group symmetries of the system [26]. Intrinsic contribution to AHE can be deduced from the first-principles calculation of the Berry curvature ($\Omega^x$, $\Omega^y$, $\Omega^z$), which acts as an emergent magnetic field in $k$-space and deflects (like Lorentz force) the carriers in the direction transverse to the applied electric field. Staggered-$\Gamma_{5g}$ phase (κ = -1, θ = $90^0$) of Mn$_3$CrN belongs to $C2/m$ magnetic space group with inversion, $C_{2[\bar{1}01]}$ and $M_{\bar{1}01}$ symmetries. Only components of AHC tensors perpendicular to $M_{\bar{1}01}$ plane (along the axis of $C_{2[\bar{1}01]}$) survives due to its constraints on the transformation properties of Berry curvature [27]. $\Omega^x$ and $\Omega^z$ are invariant under $M_{\bar{1}01}$ operation, while $\Omega^y$ is an odd function resulting in vanishing $\sigma_{xz}$ (see Extended Data Fig. 4). On the other hand, combined action of three $TM$ ($M = M_{\bar{1}01}, M_{1\bar{1}0}, M_{01\bar{1}}$) and $C_{3[111]}$ (and equivalent) symmetries of $\Gamma_{4g}$ phase (κ = 1, θ = $0^0$) of Mn$_3$CrN preserves the sign of $\Omega^\alpha$ ($\alpha = x, y, z$) throughout the Brillouin zone. From the estimated distribution of Berry curvature $\Omega^z$ at $E_F$ of N-AFM phases of Mn$_3$CrN with κ = -1 and κ = 1 for varying spin-rotation angle θ = $0^0$, $90^0$ (see Figs. 3e-h), anomalous Hall responses are expected to be dominated by the Berry curvature hotspots at X-point. Integrating $\Omega^z$ of N-AFM phases over its entire Brillouin zone leads to non-zero $\sigma_{xy}$ (see Fig. 3i) for κ = -1 (θ = $0^0$, $90^0$) and κ = 1 (θ = $0^0$, $90^0$), while the magnetic point-group symmetries of $\Gamma_{5g}$ (κ = 1, θ = $90^0$) results in vanishing $\sigma_{xy}$ (as expected when integrating $\Omega^z$ in Fig. 3h). Therefore, the first-principles DFT calculations support the observed VSC-induced concomitant magnetic-electronic-quantum geometric transition in Mn$_3$CrN.

**Epitaxial Growth and Structural Characterization of Mn$_3$CrN Films**

Structural characterization via HRXRD confirms the high crystalline, phase purity, and preferred orientation of the film (see section B and C in SM). Symmetric 2θ–ω scans reveal intense (002) and (004) Bragg reflections (see Fig. 4a) corresponding to the Mn$_3$CrN phase, with no extraneous peaks, indicating phase-pure growth along the out-of-plane [001] direction.

The narrow full-width-at-half-maximum (inset of Fig. 4a) of the rocking curve (~1.33° for the (002) peak) reflects a low mosaic spread and near single-crystalline nature of the film. Reciprocal space mapping (see Fig. 4b) around the asymmetric (113) reflection reveals a clear separation between the in-plane reciprocal lattice vector ($Q_x$) positions of the film and substrate, indicating that the Mn$_3$CrN films are fully relaxed with respect to the substrate lattice. This strain relaxation is critical, as it minimizes substrate-induced anisotropies that could otherwise distort the intrinsic magnetic ground states. Complementary pole figure analysis (see Fig. 4c) performed on the (111) reflection exhibits fourfold rotational symmetry with well-defined poles, consistent with a cube-on-cube epitaxial relationship and the absence of in-plane rotational domains or texture.

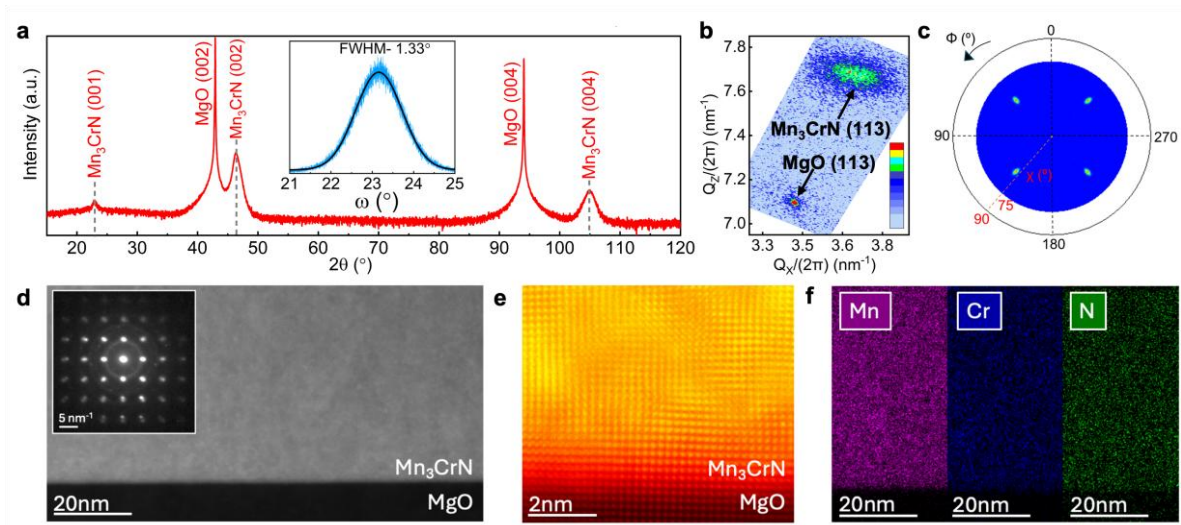

**Fig. 4| Comprehensive structural characterization of the Mn$_3$CrN film. a,** Symmetric 2θ-ω HRXRD of Mn$_3$CrN deposited on an MgO (001) substrate. The film grows with (002) orientation with a small fullwidth-at-the-half-maxima of the rocking curve of 1.33° (inset), highlighting its nominally single-crystalline nature. **b**, Reciprocal space mapping (RSM), which indicates that the film is fully relaxed, as evidenced by different $Q_x$ values of the film and the substrate. **c**, Pole figure of the (111) plane, demonstrating that the film is not textured and exhibits a single-orientation growth. **d**, Scanning transmission electron microscopy (STEM) imaging with an inset of the electron diffraction pattern, confirming cubic epitaxy. **e**, High-resolution STEM micrograph, revealing the excellent structural quality of the films. **f**, Elemental mapping of Mn, Cr, and N atoms, illustrating their uniform distribution within the film.

Cross-sectional high-angle annular dark-field scanning transmission electron microscopy (HAADF-STEM) (see Fig. 4d) provides further confirmation of the film quality. STEM images display sharp and coherent film-substrate interfaces (see Fig. 4e) as well as well-resolved lattice fringes extending across the entire film thickness, affirming the single-phase (001) [001]

Mn$_3$CrN ∥ (001) [001] MgO epitaxial growth. Selected area electron diffraction (SAED) patterns show discrete diffraction spots that match the expected cubic antiperovskite symmetry of Mn$_3$CrN. Furthermore, energy-dispersive X-ray spectroscopy (EDS) elemental mapping (see Fig. 4f) confirms the homogeneous spatial distribution of Mn, Cr, and N throughout the film, reflecting excellent stoichiometric control during deposition. X-ray photoelectron spectroscopy (XPS) (see section D in SM) further verifies the chemical composition, yielding an elemental ratio of Mn:Cr = 3.1:1.

**Discussions**

Our work uncovers a temperature-driven reversal of VSC in Mn$_3$CrN, demonstrating that chirality functions as a tunable degree of freedom capable of reorganizing the electronic topology, Berry curvature, and transport properties. The transition between κ = –1 and κ = +1 chirality states does not involve structural distortions but instead reflects a purely magnetic reconfiguration of the noncollinear antiferromagnetic texture. The chirality reversal reshapes the Fermi surface and induces a Lifshitz transition, producing a distinct evolution of XMCD response, a metal–insulator-like crossover in $\rho_{xx}$ and a pronounced peak in $\sigma_{xy}^{REM}$. The anomalous Hall response itself remains finite and sign-consistent across the entire temperature range, reflecting the robustness of Berry curvature even as the system traverses between the κ = –1 and κ = +1 chiral states. The experimental observations together with first-principles calculations, establishes that the Berry curvature landscape is governed by the VSC and the symmetry constraints of the underlying noncollinear magnetic configurations.

The VSC switching mechanism provides a conceptual parallel to symmetry-driven topological transitions in Weyl semimetals and altermagnets [48], but here the control parameter is an internal, thermodynamically accessible chiral order parameter. The multi-step evolution observed in magnetometry measurements suggests a rich magnetic energy landscape with intermediate spin-rotated configurations, characteristic of antiperovskite Kagome systems.

**Conclusion**

In summary, we uncover temperature-driven reversal of the vector spin chirality in epitaxial Mn$_3$CrN thin film, triggering a coupled magnetic–electronic–quantum geometric transition that links spin geometry to quantum transport. This emergent behaviour, captured through magnetotransport and magneto-optic measurements, and supported by first-principles theory, reveals a Lifshitz transition and Berry-curvature reconstruction accompanying a metal–insulator-like longitudinal resistivity crossover and a peak in anomalous Hall conductivity. By establishing chirality as an active control knob for quantum-geometric and electronic states, this work defines a new paradigm for tunable quantum phases in noncollinear antiferromagnets and establishes a pathway toward chirality-encoded memory, reconfigurable Hall-effect, and topological transistor devices.

**Main References**

1. Rimmler, B. H., Pal, B. & Parkin, S. S. P. Non-collinear antiferromagnetic spintronics. *Nat. Rev. Mater.* **10**, 109–127 (2025).
2. Sharma, A. K., Chatterjee, S., Yanda, P. & Felser, C. Topological quantum materials : kagome , chiral , and square-net frameworks. arXiv:2507.12410 [cond-mat.mtrl-sci] (2025).
3. Nakatsuji, S. & Arita, R. Topological Magnets: Functions Based on Berry Phase and Multipoles. *Annu. Rev. Condens. Matter Phys.* **13**, 119–142 (2022).
4. Mainz, D. J. G. Anomalous Hall effect and Spin-Orbit Torques in Noncollinear Antiferromagnetic $Mn_3Ni_{0.35}Cu_{0.65}N$ thin films, Dissertation, Johannes Gutenberg Univ. Mainz (2025).
5. Kawamura, H. Spin- and chirality-orderings of frustrated magnets--Stacked-triangular anti-ferromagnets and spin glasses. *Can. J. Phys.* **79**, 1447–1458 (2001).
6. Chen, H. Electronic chiralization as an indicator of the anomalous Hall effect in unconventional magnetic systems. *Phys. Rev. B* **106**, 024421 (2022).
7. Sharma, V., Nepal, R. & Budhani, R. C. Planar Hall effect and anisotropic magnetoresistance in thin films of the chiral antiferromagnet $Mn_3Sn$. *Phys. Rev. B* **108**, 144435 (2023).
8. Baltz, V. *et al.* Antiferromagnetic spintronics. *Rev. Mod. Phys.* **90**, 15005 (2018).
9. Guo, Z. *et al.* Spin-Polarized Antiferromagnets for Spintronics. *Adv. Mater.* **37**, 2505779 (2025).
10. Jungwirth, T., Marti, X., Wadley, P. & Wunderlich, J. Antiferromagnetic spintronics. *Nat. Nanotechnol.* **11**, 231–241 (2016).
11. Reichlova, H. *et al.* Imaging and writing magnetic domains in the non-collinear antiferromagnet $Mn_3Sn$. *Nat. Commun.* **10**, 5459 (2019).

12. Fujishiro, Y. *et al.* Giant anomalous Hall effect from spin-chirality scattering in a chiral magnet. *Nat. Commun.* **12**, 317 (2021).
13. Dally, R. L., Phelan, D., Bishop, N., Ghimire, N. J. & Lynn, J. W. Isotropic Nature of the Metallic Kagome Ferromagnet $Fe_3Sn_2$ at High Temperatures. *Crystals* **11,** 307 (2021).
14. Chen, T. *et al.* Anomalous transport due to Weyl fermions in the chiral antiferromagnets $Mn_3$X, X = Sn, Ge. *Nat. Commun.* $Mn_3$X, X = Sn, Ge. *Nat. Commun.* **12**, 572 (2021).
15. Leonov, I., Skornyakov, S. L., Anisimov, V. I. & Vollhardt, D. Correlation-Driven Topological Fermi Surface Transition in FeSe. *Phys. Rev. Lett.* **115**, 106402 (2015).
16. Varlet, A. *et al.* Tunable Fermi surface topology and Lifshitz transition in bilayer graphene. *Synth. Met.* **210**, 19–31 (2015).
17. Volovik, G. E. Exotic Lifshitz transitions in topological materials. *Phys. Usp.* **61**, 89–98 (2018).
18. Xiao, D., Chang, M. C. & Niu, Q. Berry phase effects on electronic properties. *Rev. Mod. Phys.* **82**, 1959–2007 (2010).
19. Chen, H., Niu, Q. & Macdonald, A. H. Anomalous hall effect arising from noncollinear antiferromagnetism. *Phys. Rev. Lett.* **112**, 017205 (2014).
20. Šmejkal, L., MacDonald, A. H., Sinova, J., Nakatsuji, S. & Jungwirth, T. Anomalous Hall antiferromagnets. *Nat. Rev. Mater.* **7**, 482–496 (2022).
21. Pradhan, S., Samanta, K., Saha, K. & Nandy, A. K. Vector-chirality driven topological phase transitions in noncollinear antiferromagnets and its impact on anomalous Hall effect. *Commun. Phys.* **6**, 272 (2023).
22. Keshri, A. *et al.* Unlocking Exceptional Negative Valency and Spin Reconstruction in Non-Collinear Anti-Ferromagnetic Antiperovskite $Mn_3NiN$ Film. *Adv. Funct. Mater.* **35**, 2500655 (2025).
23. Deng, S., Wang, H., He, L. & Wang, C. Magnetic structures and correlated physical properties in antiperovskites. *Microstructures* **3**, 2023044 (2023).
24. Huyen, V. T. N., Suzuki, M. T., Yamauchi, K. & Oguchi, T. Topology analysis for anomalous Hall effect in the noncollinear antiferromagnetic states of $Mn_3AN$(A = Ni, Cu, Zn, Ga, Ge, Pd, In, Sn, Ir, Pt). *Phys. Rev. B* **100**, 94426 (2019).
25. Gurung, G., Shao, D. F., Paudel, T. R. & Tsymbal, E. Y. Anomalous Hall conductivity of noncollinear magnetic antiperovskites. *Phys. Rev. Mater.* **3**, 44409 (2019).
26. Suzuki, M. T., Koretsune, T., Ochi, M. & Arita, R. Cluster multipole theory for anomalous Hall effect in antiferromagnets. *Phys. Rev. B* **95**, 094406 (2017).

27. Zhou, X. *et al.* Spin-order dependent anomalous Hall effect and magneto-optical effect in the noncollinear antiferromagnets $Mn_3X$ N with X=Ga, Zn, Ag, or Ni. *Phys. Rev. B* **99**, 104428 (2019).

28. Van Der Laan, G. Determination of spin chirality using x-ray magnetic circular dichroism. *Phys. Rev. B* **104**, 094414 (2021).

29. Mandal, A. K. *et al.* Positive Magneto-Electric Couplings in Epitaxial Multiferroic $SrMnO_3$ Thin Film. *Adv. Funct. Mater.* **35**, 2414855 (2025).

30. Piamonteze, C., Miedema, P. & De Groot, F. M. F. Accuracy of the spin sum rule in XMCD for the transition-metal L edges from manganese to copper. *Phys. Rev. B* **80**, 184410 (2009).

31. Yamamoto, S. & Matsuda, I. Measurement of the resonant magneto-optical Kerr effect using a free electron laser. *Appl. Sci.* **7**, 662 (2017).

32. Banik, S. Electrical Detection Of Spin Hall Effect By Magneto Optical Kerr Effect. (NISER Bhubaneswar, 2023).

33. Teixeira, J. M. *et al.* Versatile, high sensitivity, and automatized angular dependent vectorial Kerr magnetometer for the analysis of nanostructured materials. *Rev. Sci. Instrum.* **82,** 043902 (2011).

34. Pan, H. *et al.* Orthogonal Geometry of Magneto-Optical Kerr Effect Enabled by Magnetization Multipole of Berry Curvature. arXiv:2412.09857 [cond-mat.mtrl-sci] (2025).

35. Higo, T. *et al.* Large magneto-optical Kerr effect and imaging of magnetic octupole domains in an antiferromagnetic metal. *Nat. Photonics* **12**, 73–78 (2018).

36. Öztürk, M. Coercivity Enhancement and the Analysis of Asymmetric Loops in a Perpendicularly Magnetized Thin Film. *J. Supercond. Nov. Magn.* **33**, 3097–3105 (2020).

37. Biswas, B. *et al.* Magnetic Stress-Driven Metal-Insulator Transition in Strongly Correlated Antiferromagnetic CrN. *Phys. Rev. Lett.* **131**, 126302 (2023).

38. Lee, D. *et al.* Isostructural metal-insulator transition in $VO_2$. *Science* **362**, 1037–1040 (2018).

39. Mott, N. F. Metal-Insulator Transitions. *Pure Appl. Chem.* **52**, 65–72 (1980).

40. Siegrist, T. *et al.* Disorder-induced localization in crystalline phase-change materials. *Nat. Mater.* **10**, 202–208 (2011).

41. Liu, E. *et al.* Giant anomalous Hall effect in a ferromagnetic kagome-lattice semimetal. *Nat. Phys.* **14**, 1125–1131 (2018).

42. Roychowdhury, S. *et al.* Anomalous Hall Conductivity and Nernst Effect of the Ideal Weyl Semimetallic Ferromagnet EuCd$_2$As$_2$. *Adv. Sci.* **10**, 2207121 (2023).

43. Nagaosa, N., Sinova, J., Onoda, S., MacDonald, A. H. & Ong, N. P. Anomalous Hall effect. *Rev. Mod. Phys.* **82**, 1539–1592 (2010).

44. Swamy, G. V., Rakshit, R. K., Pant, R. P. & Basheed, G. A. Origin of 'in-plane' and 'out-of-plane' magnetic anisotropies in as-deposited and annealed CoFeB ferromagnetic thin films. *J. Appl. Phys.* **117**, 17A312 (2015).

45. Rijks, T. G. S. M., Lenczowski, S. & Coehoorn, R. In-plane and out-of-plane anisotropic magnetoresistance in thin films. *Phys. Rev. B* **56**, 362–366 (1997).

46. Segall, M. D. *et al.* First-principles simulation: Ideas, illustrations and the CASTEP code. *J. Phys. Condens. Matter* **14**, 2717–2744 (2002).

47. P. Geerlings, F. De Proft, W. Langenaeker, Conceptual Density Functional Theory. *Chem. Rev.* **103***, 1793–1873* (2003).

48. Cheong, S. W. & Huang, F. T. Altermagnetism with non-collinear spins. *npj Quantum Mater.* **9**, 13 (2024).

# Methods

**Film Growth:** Mn$_3$CrN thin films are deposited on single-crystal (001) MgO and SrTiO$_3$ (STO) substrates (1 cm × 1 cm) using an ultrahigh vacuum (UHV) reactive DC magnetron sputtering (PVD Products Inc.) operating at a base pressure of $1 \times 10^{-9}$ Torr. High-purity Mn and Cr targets (diameter: 2 inches; thickness: 0.25 inches) were co-sputtered at 12W and 8 W, respectively, leading to the stoichiometric deposition of the Mn$_3$CrN. Prior to deposition, substrates were ultrasonically cleaned in acetone and methanol for 10 minutes each to remove organic contaminants. During the deposition, a total working gas pressure of 5 mTorr was maintained using a gas mixture of Ar and N$_2$ in a 9:2 standard cubic centimeters per minute (sccm) flow ratio. These optimized parameters enabled the growth of high-quality epitaxial Mn$_3$CrN thin films with excellent phase purity, surface morphology, and crystallographic alignment.

**Electrical Measurements**: The electronic phase transition and anomalous Hall effect (AHE) in Mn$_3$CrN thin films are investigated with temperature- and field-dependent electrical measurements. All transport measurements were carried out using a Physical Property Measurement System (PPMS, Quantum Design) in the temperature range of 2-375K with a maximum magnetic field (H) of $\pm$9T, employing both standard four-probe and Van der Pauw geometries. For temperature sweeps, a consistent step size of 2 K was used across the entire measurement range of 2-375K to capture fine details of the resistivity evolution. For magnetic field sweeps, a consistent step size of 500 Oe was maintained across entire measurement range of $\pm$9T to capture fine details of the resistivity evolution.

**Magnetic Measurements**: The magnetic measurements were performed using a superconducting quantum interference devices (SQUID) magnetometer (MPMS-3, Quantum Design) in the temperature range of 2-400K with a maximum magnetic field (H) of $\pm$7T. Before each magnetic measurement, the superconducting magnet was reset to eliminate any residual magnetic field within the coil. Subsequently, the sample was demagnetized at 300 K, well above the transition temperature using an oscillating magnetic field under a standardized demagnetization protocol.

**XAS & XMCD measurements**: Element-specific X-ray absorption spectroscopy (XAS) and X-ray magnetic circular dichroism (XMCD) measurements were carried out at the BOREAS beamline of the ALBA Synchrotron (Barcelona, Spain). Spectra were recorded at the Mn and Cr $L_{3,2}$ edges and the N and O $K$ edges using the surface-sensitive total electron yield (TEY) mode, by measuring the sample drain current. Measurements were performed at 3.5 K, 20 K, 80 K, 180 K and 300 K, where the X-ray beam impinged on the film surface at an angle of 55° under an applied magnetic field of 6 T. Angle-dependent XMCD measurements were further conducted at incidence angles of 20°, 55°, and 90° at 3.5 K under the same field conditions. All experiments were performed under a base pressure of $8 \times 10^{-9}$ Torr.

Circularly polarized X-rays with ~90% polarization were used in the XMCD measurements. The XMCD was obtained by taking the difference of the XAS spectra of right circularly polarized light ($\mu^+$) and left circularly polarized light ($\mu^-$), i.e., $\mu^+ - \mu^-$, by flipping the X-ray helicity at a fixed magnetic field of 6 T. The normalization of the spectra was performed so that the $L_3$ pre-edge spectral region was set to zero. The sum rules (see Extended Data Fig. 1) are used to calculate the orbital magnetic moment ($l_Z$) and the spin magnetic moment ($s_Z$) of the Mn$_3$CrN thin film. According to the sum rules, the $s_Z$ and $l_Z$ are given as

$$s_Z = -N^d \left(\frac{6p-4q}{r}\right) \tag{S1}$$

$$l_Z = -\frac{4}{3} N^d \left(\frac{q}{r}\right) \tag{S2}$$

Where, $p = \int_{L_3} (\mu^+ - \mu^-) \, dE$; $q = \int_{L_{2,3}} (\mu^+ - \mu^-) \, dE$; $r = \int_{L_{2,3}} (\mu^+ + \mu^-) \, dE$; and $N^d$ is the number of holes in the 3$d$ orbital, which we considered as 5 for Mn. An arctangent step background was fitted and subtracted from both the $\mu^+$ and $\mu^-$ spectra to remove the nonmagnetic contribution. To ensure robustness, the XMCD spectra were also analysed using alternative background models (linear and step functions), but no significant differences were found in the extracted moment values.

**MOKE Measurement:** A Kerr microscopy–based magneto-optic Kerr effect (MOKE) setup is used to determine the magneto-optic properties of Mn$_3$CrN thin films. Both longitudinal and polar geometries were employed to probe the in-plane and out-of-plane magnetization components, respectively. In the longitudinal geometry, the incident light was directed at a small angle relative to the film surface, making the signal sensitive to the magnetization parallel to the applied field. In the polar geometry, normal incidence was used, which is sensitive to the out-of-plane component of magnetization. The Kerr rotation was recorded as a function of the applied magnetic field to extract hysteresis loops, while Kerr microscopy enabled imaging of magnetic domain evolution under field, providing complementary insight into magnetization reversal processes.

**High-resolution X-ray Diffraction:** High-resolution X-ray diffraction (HRXRD) measurements are performed to determine the crystal structure of $Mn_3CrN$ films using the Rigaku SmartLab X-ray diffractometer with Cu-Kα source of wavelength 1.54 Å. A 4.5 kW rotating anode X-ray generator was used in a parallel beam optics geometry with a multi-layer X-ray mirror, Germanium (220) 2-bounce channel cut monochromator and 2-bounce analyzer are used for 2θ-ω scan, rocking curve, pole figure and reciprocal space mapping (RSM). RSM measurements of asymmetric (113) planes were performed with a HyPix-400 vertical detector in 0D continuous exposure mode. Before every measurement, optical and sample alignment were performed. Pole figure measurements of the (111) plane with respect to the (002) growth plane are carried out in the Parallel Beam (PB) optics mode by varying the χ angle from 0 to 75°, fixing the 2θ and ω of the (111) plane. The in-plane and out-of-plane lattice parameters were obtained from RSM data by considering the substrate MgO's lattice constant to 4.21 Å.

**High-resolution Transmission Electron Microscopy Imaging:** Cross-sectional high-angle annular dark-field scanning transmission electron microscopy (HAADF-STEM) imaging and energy-dispersive X-ray spectroscopy (EDS) mapping were performed using an image- and probe-corrected monochromated FEI Themis-Z transmission electron microscopy operated at 300 kV equipped with a high brightness XFEG source and Super-X EDS detector for ultra-high-count rates. The spatial resolution of the STEM mode is 0.7 Å. EDS maps used for atomic concentration quantification is recorded with well above 1 million count rates along with the background correction using k-factor analysis and absorption correction.

**Computational Details:** First-principles density functional theory (DFT) calculations were performed for the non-collinear antiferromagnetic (coplanar) phases of $Mn_3CrN$ using the VASP package [49, 50]. The projector augmented wave (PAW) method was employed to describe valence electron–ion core interactions [51], with a plane-wave energy cutoff of 600 eV. Exchange–correlation effects were treated within the generalized gradient approximation (GGA) using the Perdew–Burke–Ernzerhof (PBE) functional [52]. Structural optimization of the NC-AFM phase was carried out within the DFT+U framework, following the approach of Liechtenstein *et al.* [53], with an on-site U value of 2 eV applied to Mn *d* states to reproduce the experimental lattice parameters and local Mn magnetic moments (see Extended Data Fig. 3 for details). Brillouin zone integrations were performed using a Γ-centered 12 × 12 × 12 $k$-mesh for the cubic Mn$_3$CrN lattice. Spin–orbit coupling (SOC) was included in all electronic structure calculations.

Maximally localized Wannier functions (MLWFs) were constructed using the Wannier90 package [54], with $s, p, d$ orbitals projected onto Mn and Cr atoms and $s, p$ orbitals projected onto N. The resulting tight-binding Hamiltonian was employed to evaluate Berry curvature components and the intrinsic anomalous Hall conductivity tensor on a 112 × 112 × 112 FFT grid using the WannierBerri code [55]

$$\sigma_{\alpha\beta} = -\frac{e^2}{\hbar}\epsilon_{\alpha\beta\gamma}\int_{BZ}\sum_n \frac{d^3\mathbf{k}}{(2\pi)^3}\Omega_n^\gamma(\mathbf{k})f_n(\mathbf{k}) \quad (S3)$$

where $f_n(\mathbf{k})$ is the Fermi-Dirac distribution function, $\epsilon_{\alpha\beta\gamma}$ is the anti-symmetric tensor relating the transverse Hall conductivity $\sigma_{\alpha\beta}$ and $\gamma^{\text{th}}$ component of band-resolved ($n$) Berry curvature $\Omega_n^\gamma(\mathbf{k})$, calculated as:

$$\Omega_n^\gamma(\mathbf{k}) = -2\text{Im}\sum_{m\neq n}\frac{\langle\psi_{n\mathbf{k}}|\hat{v}_\alpha|\psi_{m\mathbf{k}}\rangle\langle\psi_{m\mathbf{k}}|\hat{v}_\beta|\psi_{n\mathbf{k}}\rangle}{[\varepsilon_m(\mathbf{k})-\varepsilon_n(\mathbf{k})]^2} \quad (S4)$$

Here, $\psi_{n\mathbf{k}}$ is the Bloch function with a band index $n$ and momentum $\mathbf{k}$, $\hat{v}_\alpha$ is the velocity operator along the $\alpha$ direction and $\varepsilon_n(\mathbf{k})$ the energy eigenvalues of band $n$. We confirmed the convergence of anomalous Hall conductivities for varying FFT grids in WannierBerri. We also verified the consistency of our calculated $\Omega_n^\gamma(\mathbf{k})$ and $\sigma_{\alpha\beta}$ with Wanniertools [56] and postW90 code respectively.

**Acknowledgements:** A.D. and B.S. acknowledge the International Centre for Materials Science (ICMS) and Sheikh Saqr Laboratory (SSL) of Jawaharlal Nehru Centre for Advanced Scientific Research (JNCASR) for support. B.S. acknowledges the Anusandhan National Research Foundation (ANRF) in India for a Core Research Grant No. CRG/2023/007061 for partial financial support. A.D., A. B., and B.S. acknowledge the *SAMat* Research Facilities, JNCASR, Bengaluru. R.B. acknowledges SASTRA Deemed University, Thanjavur for financial assistance through a fellowship. M.G. and A.I.K.P. acknowledge the facilities of Sydney Microscopy and Microanalysis at the University of Sydney. T.M., M.B. and R.M. acknowledge IISERTVM's magnetic measurement facilities.

**Author contributions:** A.D. and B.S. conceived this project. A.D. deposited the thin films and performed structural characterizations. A.D., A.B. S.B. performed magnetotransport measurements. M.B., R.M. and T. M. performed magnetic measurements, S.C., J.K.D., W. L. and M.V. performed XMCD measurements. A.I.K.P. performed the TEM sample preparation and M.G. performed the TEM imaging and analysis. R. B. and U.V.W. performed first-principles theoretical modelling. S.R., U.V.W, provided valuable insight into interpretations of the experimental results. All authors discussed and contributed to the preparation of the manuscript.

**Conflict of interest:** There are no conflicts of interest to declare.

**Data availability statement:** All data required to evaluate the conclusions in the manuscript are available in the main text or the supplementary materials.

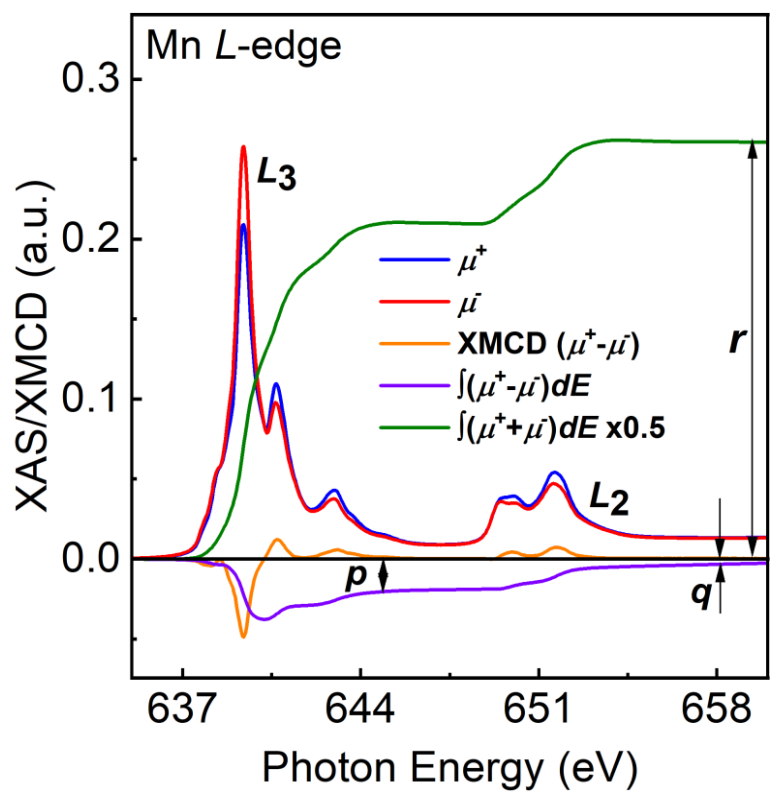

**Extended Data Fig. 1 | Sum rule analysis of Mn $L_{3,2}$-edges XAS and XMCD.** XAS spectra for parallel (μ+, blue) and antiparallel (μ-, red) polarization, the XMCD difference (orange), and integrated XAS (green) and XMCD (purple) signals are shown. Integrated XMCD confirms a robust magnetic moment for Mn, enabling quantitative extraction of spin and orbital contributions by sum rule formalism.

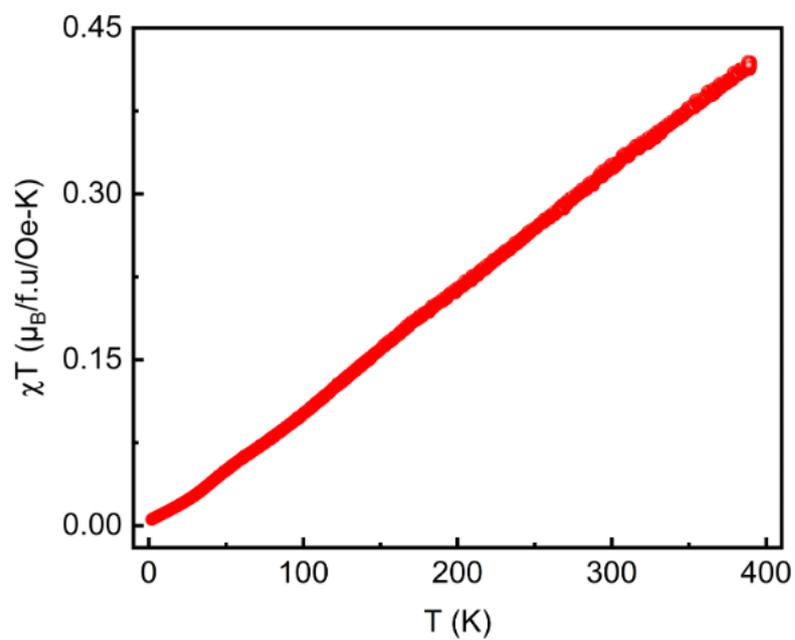

**Extended Data Fig. 2 |** Temperature evolution of χT, showing a linear increase with temperature, consistent with the antiferromagnetic Curie–Weiss behavior ($\chi T = C/(1 + T_N/T)$), where $T_N$ is the Neel temperature. [57]

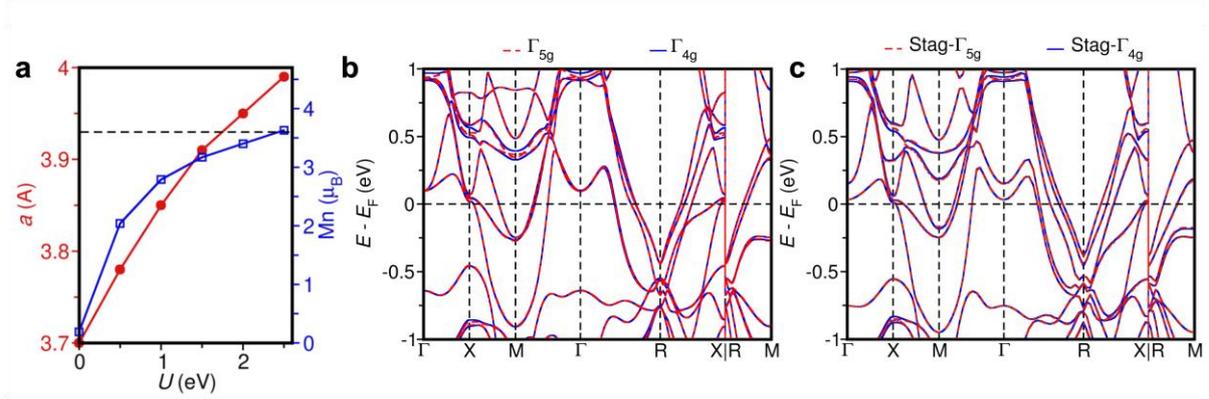

**Extended Data Fig. 3 | Structural parameter, magnetic moment at Mn and electronic properties of Mn$_3$CrN from first-principles calculations. a**, Optimized cubic lattice parameter and local Mn magnetic moments as a function of onsite Coulomb energy $U$. The dashed lines mark the experimental lattice constant and the magnetic moment measured at 2 K. **b**, Electronic structures of the $\Gamma_{4g}$ (blue) and $\Gamma_{5g}$ (red) spin configurations, showing nearly identical dispersion of the frontier states. **c**, Electronic structures of the staggered-$\Gamma_{4g}$ (blue) and staggered-$\Gamma_{5g}$ (red) phases, likewise exhibiting indistinguishable frontier bands, confirming their electronic energy degeneracy.

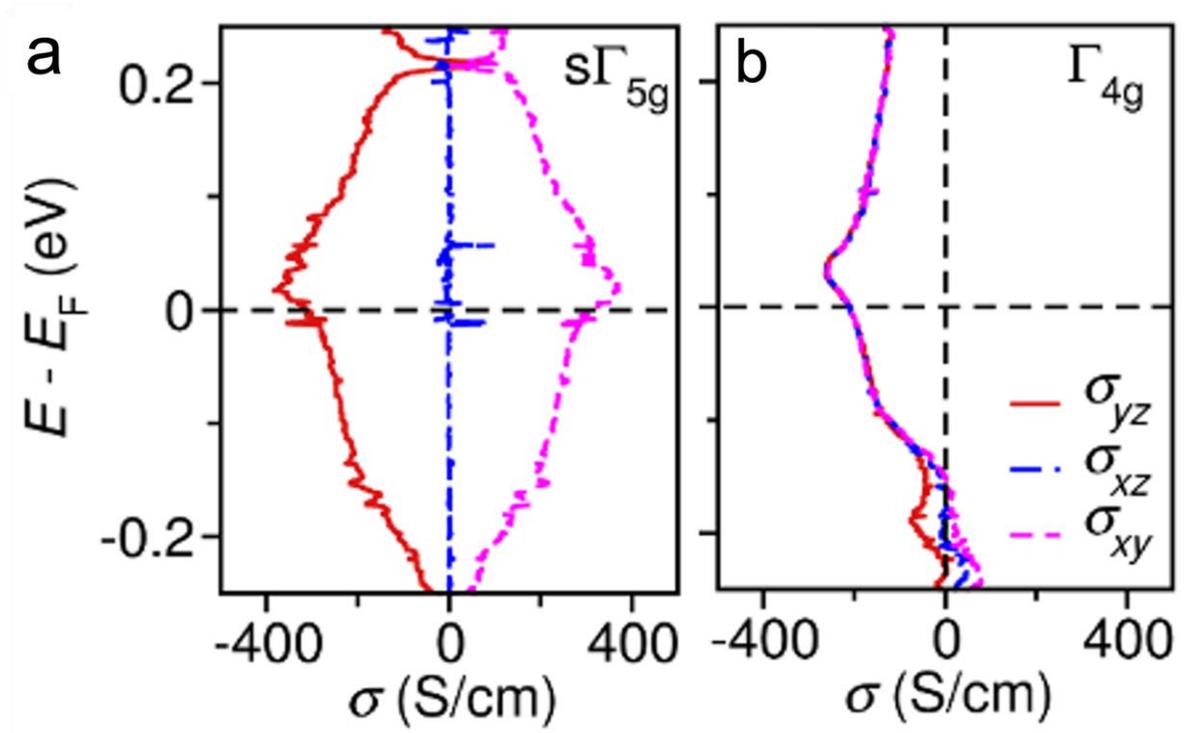

**Extended Data Fig. 4 |** Energy-dependent anomalous Hall conductivity (σ) of $Mn_3CrN$ for **a**, staggered-$\Gamma_{5g}$ and **b**, $\Gamma_{4g}$ states. The s$\Gamma_{5g}$ phase exhibits nearly vanishing $\sigma_{xz}$, whereas the $\Gamma_{4g}$ phase hosts finite $\sigma_{xz}$, evidencing symmetry-controlled anisotropy in Berry curvature and Hall response.

**Extended Data Table 1. Energetics between different chirality configuration of $\Gamma_{4g}$ and $\Gamma_{5g}$**

|  | Stag $\Gamma_{5g}$ - $\Gamma_{5g}$ | Stag $\Gamma_{4g}$ - $\Gamma_{4g}$ | Stag $\Gamma_{4g}$ - Stag $\Gamma_{5g}$ | $\Gamma_{4g}$ - $\Gamma_{5g}$ |
|---|---|---|---|---|
| ΔE (meV/f.u.) | -87 | -85 | 0.45 | -0.67 |

**References:**

49. Hobbs, D., Kresse, G. & Hafner, J. Fully unconstrained noncollinear magnetism within the projector augmented-wave method. *Phys. Rev. B* **62**, 11556–11570 (2000).

50. Vargas-Hernández, R. A. Bayesian Optimization for Calibrating and Selecting Hybrid-Density Functional Models. *J. Phys. Chem. A* **124**, 4053–4061 (2020).

51. Joubert, D. From ultrasoft pseudopotentials to the projector augmented-wave method. *Phys. Rev. B* **59**, 1758–1775 (1999).

52. Li, K., Luo, L., Zhang, Y., Li, W. & Hou, Y. Tunable Luminescence Contrast in Photochromic Ceramics (1 - x)Na$_{0.5}$Bi$_{0.5}$TiO$_3$ - xNa$_{0.5}$K$_{0.5}$NbO$_3$:0.002Er by an Electric Field Poling. *ACS Appl. Mater. Interfaces* **10**, 41525–41534 (2018).

53. Liechtenstein, A. I., Anisimov, V. I. & Zaanen, J. Density-functional theory and strong interactions: Orbital ordering in Mott-Hubbard insulators. *Phys. Rev. B* **52**, 5467–5471 (1995).

54. Pizzi, G. *et al.* Wannier90 as a community code: New features and applications. *J. Phys. Condens. Matter* **32**, 165902 (2020).

55. Tsirkin, S. S. High performance Wannier interpolation of Berry curvature and related quantities with WannierBerri code. *npj Comput. Mater.* **7**, 33 (2021).

56. Wu, Q. S., Zhang, S. N., Song, H. F., Troyer, M. & Soluyanov, A. A. WannierTools: An open-source software package for novel topological materials. *Comput. Phys. Commun.* **224**, 405–416 (2018).

57. Singh, J., Panda, S. K. & Singh, A. K. Recent developments in supramolecular complexes of azabenzenes containing one to four N atoms: synthetic strategies, structures, and magnetic properties. *RSC Adv.* **12**, 18945–18972 (2022).

# Supplementary Materials

**Contents:**

**A:** Crystal Structure and Kagome Plane in $Mn_3CrN$ Thin Films

**B:** Structural Characterization of $Mn_3CrN$ Thin Films on STO and MgO

**C:** 2D Reciprocal Space Mapping (RSM) Analysis of $Mn_3CrN$ Thin Films

**D:** X-ray Photoelectron Spectroscopy (XPS) Analysis of Mn, Cr, and N Core Levels

**E:** Element-Specific X-ray Absorption Spectroscopy Measurements

**F:** Temperature-dependent resistivity fitting of $Mn_3CrN$ film

**G:** Intrinsic origin of the anomalous Hall effect revealed by scaling analysis

**H:** Anomalous Hall Effect in $Mn_3CrN$ Thin Films on STO Substrate

**I:** Anomalous Hall Coercivity in $Mn_3CrN$/MgO

**J:** First principles DFT study of $Mn_3CrN$

**K:** Energetic penalty of out-of-plane spin canting in $Mn_3CrN$

**L:** Spin Rotation Continuum Between $\Gamma_{4g}$ and $\Gamma_{5g}$

**M:** Spin Rotation Continuum Between staggered-$\Gamma_{4g}$ and staggered-$\Gamma_{5g}$

**Figure S1 to S14**

**References**

## A. Crystal Structure and Kagome Plane in Mn$_3$CrN Thin Films

Mn$_3$CrN belongs to the family of antiperovskite nitrides, which adopt a cubic perovskite-derived structure in which the conventional roles of anions and cations are reversed. It crystallizes in the cubic space group $Pm\bar{3}m$ (No. 221), with chromium atoms at the cube corners (0, 0, 0), nitrogen atoms located at the body-centered position (½, ½, ½), and manganese atoms distributed over the face-centered positions (½, ½, 0), (½, 0, ½), and (0, ½, ½), as illustrated in Fig. S1a. This arrangement forms a three-dimensional network of corner-sharing octahedra, with nitrogen at the centers and Mn–Cr polyhedra providing the structural backbone of the lattice.

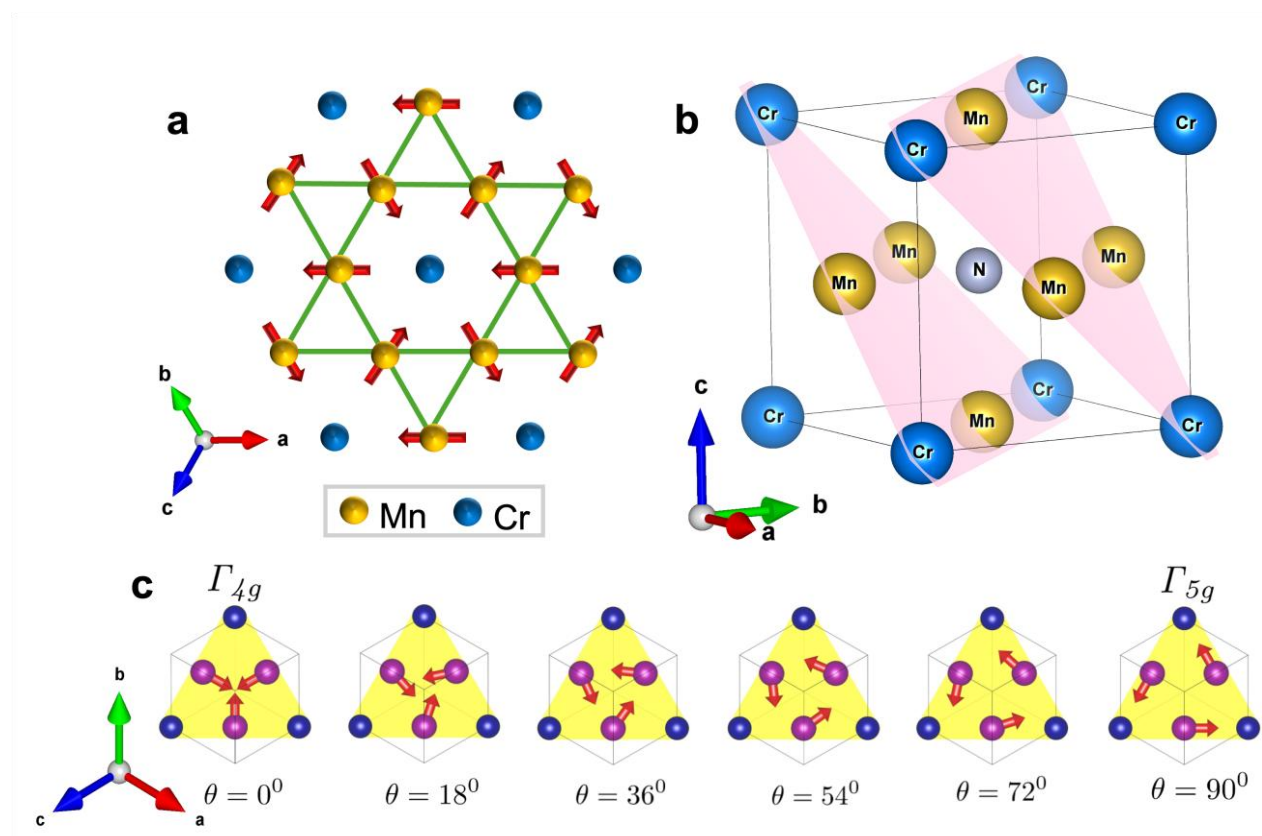

**Fig. S1 | Crystal and Kagome lattice structure of Mn$_3$CrN. a**, Kagome network of Mn$_3$CrN viewed along the [111] direction. Mn atoms (yellow) form a corner-sharing triangular lattice with coplanar non-collinear spin arrangement (red arrows), while Cr atoms (blue) are positioned between the Mn Kagome layers. **b**, Mn (yellow) and Cr (blue) atoms in the (111) plane form Kagome plane. N atoms (grey) occupy sites between these planes. **c**, Systematic evolution from $\Gamma_{4g}$ to $\Gamma_{5g}$ through consecutive spin rotations within the Kagome plane.

From a crystallographic perspective, the most significant feature is the arrangement of Mn atoms. When viewed along the [111] axis, the Mn sublattice maps onto a Kagome-type network (see Fig. S1a), consisting of corner-sharing triangles stacked in alternating layers. The $\Gamma_{4g}$ and $\Gamma_{5g}$ spin configurations, along with the intermediate spin rotations between them, are illustrated in Fig. S1c.

The lattice parameter of $Mn_3CrN$ thin films, determined from high-resolution XRD measurements, is $a$ = 3.93 Å at room temperature. Notably, $Mn_3CrN$ preserves its cubic symmetry throughout the entire temperature range investigated, showing no evidence of structural distortions (see Fig. S2), such as the tetragonal or orthorhombic transitions commonly observed in other correlated antiperovskites.[1] This robustness of the cubic phase underscores the dominant influence of electronic and magnetic interactions over lattice instabilities.

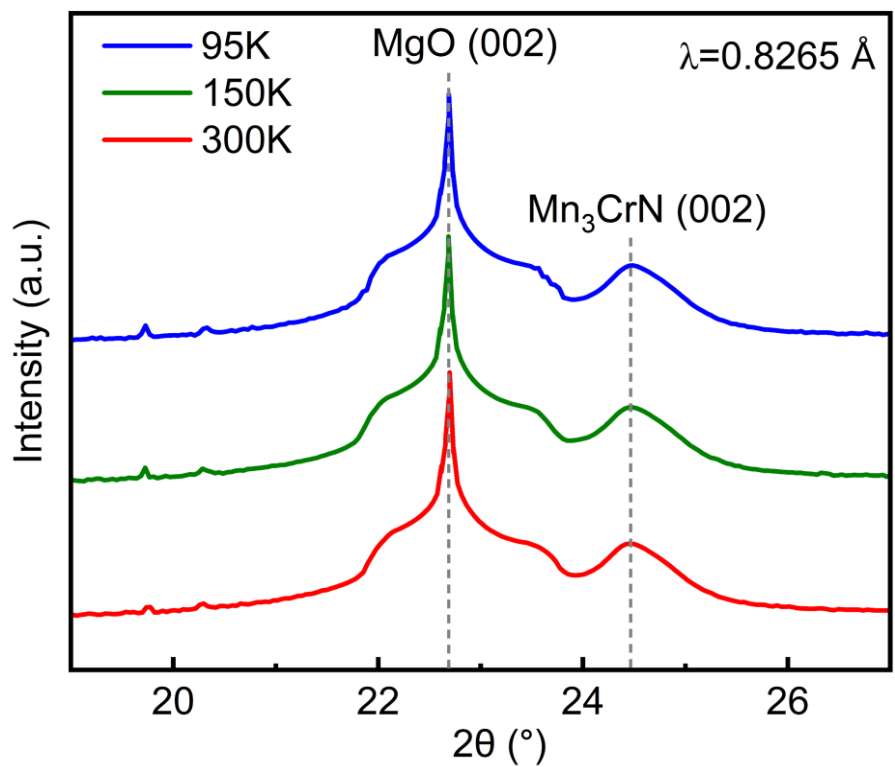

**Fig. S2** | Temperature-dependent XRD 2θ-ω scans of $Mn_3CrN$ thin films on MgO (001) substrate, measured at 95 K, 150 K, and 300 K using synchrotron radiation (λ = 0.8265 Å). The $Mn_3CrN$ (002) reflection shows no discernible peak shift with temperature, confirming the absence of structural transitions and the persistence of the cubic antiperovskite phase across the investigated range.

At the local coordination level, each Mn site is surrounded by two nitrogen and four chromium atoms, forming a slightly distorted octahedral cage. Conversely, the central nitrogen atom is octahedrally coordinated by six Mn atoms, creating a symmetric N–Mn$_6$ octahedron. This geometry fosters strong Mn–N hybridization, confirmed by spectroscopic studies, and plays a crucial role in stabilizing itinerant magnetism within the system.

B. **Structural Characterization of Mn$_3$CrN Thin Films on STO and MgO**

To assess the crystal structure, epitaxy, and microstructural properties of Mn$_3$CrN thin films, grazing-incidence in-plane XRD, X-ray reflectivity (XRR), and pole figure measurements on films grown on MgO and SrTiO$_3$ (STO) substrates are perfromed. Fig. S3a shows the 2θ–ω

XRD scan of Mn₃CrN thin films deposited on STO (001), alongside the bare substrate reference. The Mn₃CrN (002) reflection nearly overlaps with the STO (002) peak, reflecting the close lattice parameter match between film and substrate and making the peaks difficult to resolve separately. This near-coincidence indicates epitaxial growth with minimal out-of-plane strain, demonstrating excellent crystallographic compatibility. The absence of secondary reflections further confirms phase purity and highly oriented (001) growth.

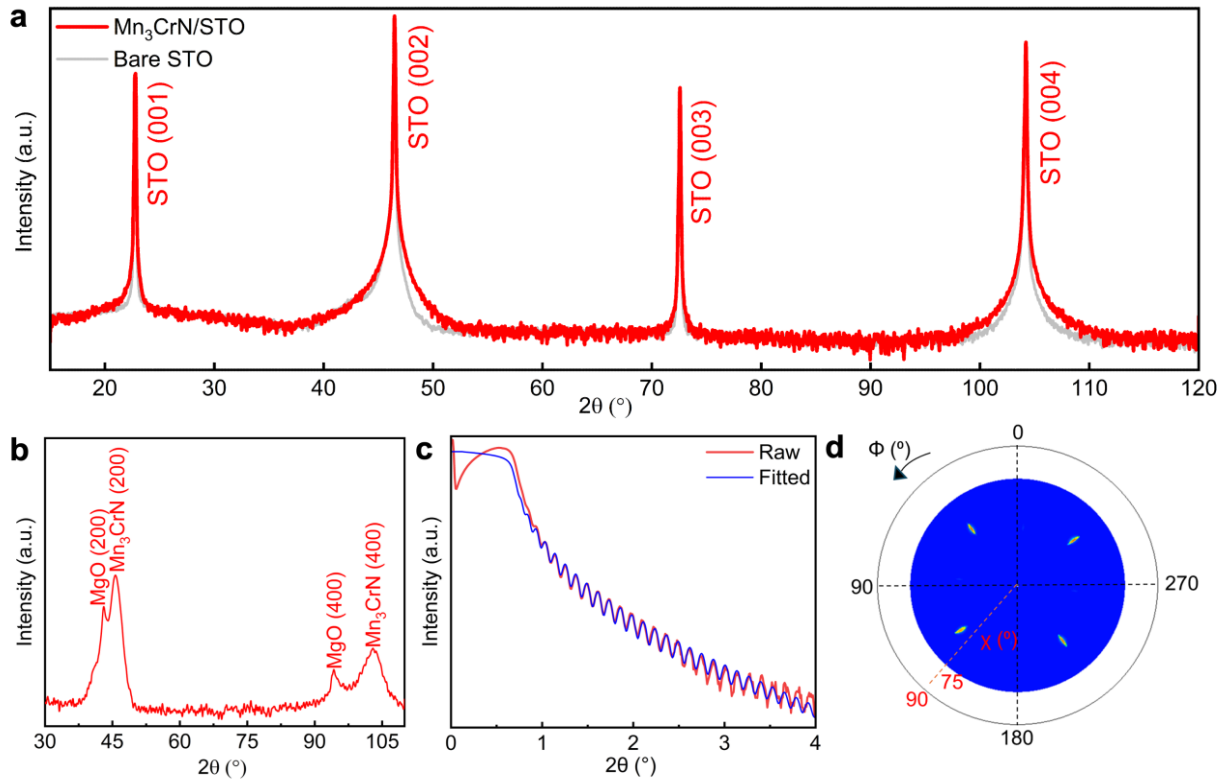

**Fig. S3 | Structural Characterization of Mn₃CrN Thin Films on STO and MgO. a**, 2θ-ω XRD scan of Mn₃CrN thin film on SrTiO₃ (001) substrate (red) compared with bare STO (gray). The Mn₃CrN (002) peak is nearly merged with the STO (002) reflection due to the closely matched lattice constants, indicating strain-free epitaxial growth and high crystalline quality, **b**, In-plane grazing incidence XRD scan for Mn₃CrN grown on MgO, showing the Mn₃CrN (200)/(020) peak at the same position as the (002) out-of-plane peak, confirming a cubic crystal structure, **c**, X-ray reflectivity (XRR) of Mn₃CrN/MgO film with data (red) and fit (blue) using the Parratt formalism. The extracted film parameters include a thickness of ~70 nm, density of ~4.28 g/cm³, and surface roughness of ~0.396 nm, **d**, Pole figure of the (111) reflection from Mn₃CrN/STO film showing fourfold symmetry and absence of out-of-plane texture, confirming epitaxial growth on the cubic substrate.

Fig. S3b shows the in-plane grazing incidence XRD scan measured on the Mn₃CrN/MgO thin film. The observed Mn₃CrN (200)/(020) peak appears at the same 2θ position as the Mn₃CrN

(002) peak seen in out-of-plane scans, confirming the cubic symmetry of the antiperovskite structure.

Fig. S3c displays the XRR measurement of the Mn$_3$CrN/MgO film, fitted using the Parratt algorithm. The fit yields a film thickness of approximately 70 nm, an average surface roughness of ~0.396 nm, and a density of ~4.28 g/cm³, which closely matches the theoretical value for stoichiometric Mn$_3$CrN. This indicates that the film has a smooth surface and compact growth morphology.

The (111) pole figure in Fig. S3d was obtained from Mn$_3$CrN/STO films. The fourfold symmetry observed in the intensity distribution confirms epitaxial cube-on-cube alignment with the cubic STO substrate. The absence of a preferred χ-orientation further suggests that the (111) planes are not textured, and the film maintains uniform in-plane symmetry without twinning.

Together, these results validate that the Mn$_3$CrN thin films possess high crystallinity, epitaxial alignment, and preserve their intended cubic antiperovskite structure on both MgO and STO substrates.

**C. 2D Reciprocal Space Mapping (RSM) Analysis of Mn$_3$CrN Thin Films**

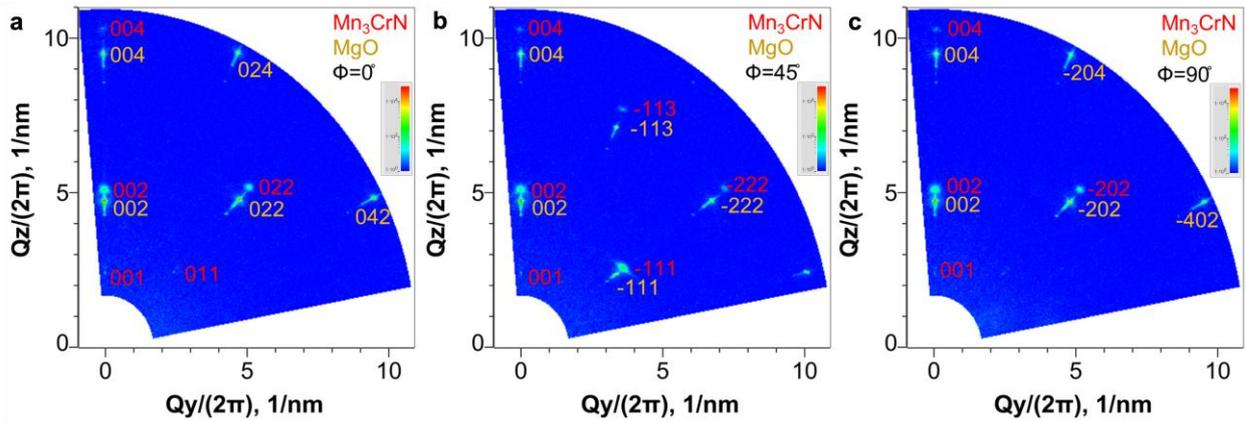

**Fig. S4 | 2D Reciprocal Space Mapping (RSM) Analysis of Mn₃CrN Thin Films. a**, **b**, **c** shows two-dimensional reciprocal space maps (2D-RSMs) of Mn$_3$CrN thin films on MgO (001) at φ = 0°, 45°, and 90° respectively, showing symmetric and off-axis Bragg reflections. The presence of well-defined, symmetry-consistent peaks confirms the cubic crystal structure and single-domain cube-on-cube epitaxial growth.

To assess the in-plane crystallographic symmetry and epitaxial quality of the Mn$_3$CrN thin films, φ-dependent two-dimensional reciprocal space mappings (2D-RSMs) were performed at azimuthal angles φ = 0°, 45°, and 90°, as shown in Fig. S4(a–c).[2] At φ = 0°, symmetric Bragg reflections such as (001), (002), (004), and off-axis peaks including (011), (022), and (042) from Mn$_3$CrN are observed, alongside their MgO substrate counterparts. This configuration primarily probes the out-of-plane [001] axis and confirms the vertical orientation and periodicity of the film. At φ = 45°, the RSM reveals a new set of reflections—specifically (111), (113), and (222)—which lie along diagonal directions and are characteristic of a cubic crystal structure viewed along intermediate axes. These peaks affirm the presence of three-dimensional ordering and further validate the FCC-like symmetry of the antiperovskite Mn$_3$CrN lattice. At φ = 90°, a different but symmetry-equivalent set of reflections emerges, including (002), (004), (202), and (402), which are consistent with in-plane rotation by 90°, confirming the fourfold rotational symmetry of the cubic phase. The consistency of reflection positions and intensities across φ angles demonstrates the absence of in-plane twinning or misoriented grains and reinforces the cube-on-cube epitaxial relationship between the Mn$_3$CrN film and the MgO (001) substrate. These results collectively confirm the single-domain, high-quality epitaxial nature of the Mn$_3$CrN films and align with other structural characterizations such as HRXRD and pole figure analysis.

**D. X-ray Photoelectron Spectroscopy (XPS) Analysis of Mn, Cr, and N Core Levels**

High-resolution XPS was performed to investigate the chemical states of Mn, Cr, and N in $Mn_3CrN$ thin films, and the core-level spectra were deconvoluted to resolve bonding environments. The Mn-*2p* region at Fig. S5a displays a clear spin–orbit doublet with components centered at Mn-$2p_{3/2}$ (637.8, 638.9, and 640.9 eV) and Mn-$2p_{1/2}$ (649.0, 649.69, and 650.88 eV), with an energy separation of 11.35 eV and a consistent area ratio of 2:1, as expected for Mn-*2p* levels. The peaks at 637.8 eV and 649.0 eV correspond to Mn–N bonding in the Mn₃CrN lattice, while the features at 638.9 eV and 649.69 eV arise from Mn–O–N environments associated with partial surface oxidation or coordination disorder, and the higher-binding-energy peaks at 640.9 eV and 650.88 eV are attributed to $Mn^{2+}$–O species such as MnO, reflecting the formation of a thin surface oxide. No metallic Mn signal was detected, confirming that manganese is present exclusively in nitride and oxidized forms.

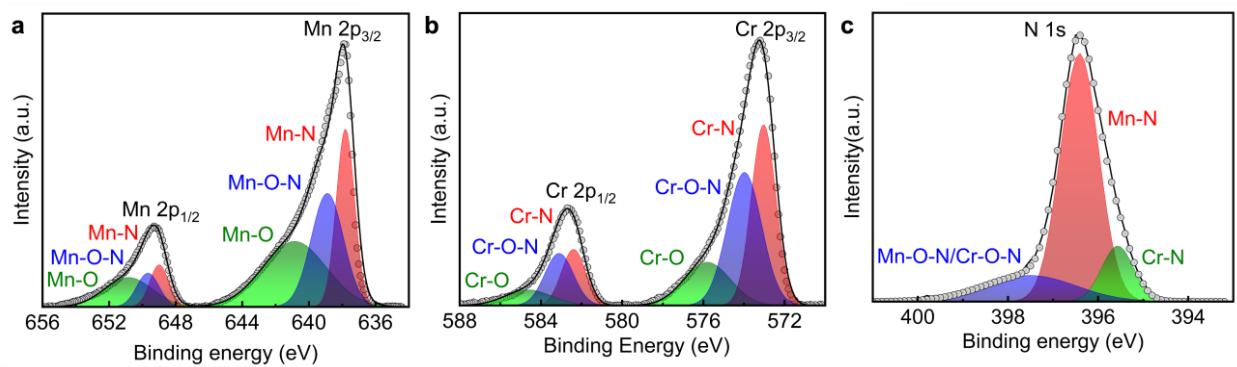

**Fig. S5 | High-resolution XPS spectra of Mn₃CrN thin films. a**, Mn 2p showing spin–orbit splitting with components from Mn–N, Mn–O–N, and Mn–O species, **b**, Cr 2p with resolved Cr–N, Cr–O–N, and Cr–O contributions, and **c**, N 1s dominated by Mn–N and Cr–N bonding with a minor Mn–O–N/Cr–O–N feature. The absence of metallic Mn or Cr signals confirms the chemically stable rocksalt Mn₃CrN phase with only slight surface oxidation typical of nitride films.

The Cr-*2p* spectrum at Fig. S5b likewise shows a spin–orbit doublet with peaks at Cr-2p$_{3/2}$ (573.02, 574.00, and 575.95 eV) and Cr-2p$_{1/2}$ (582.38, 583.14, and 584.83 eV), with a measured splitting of 9.42 eV and a 2:1 area ratio. The components at 573.02 and 582.38 eV are characteristic of Cr–N bonding within the rocksalt Mn₃CrN lattice, while the peaks at 574.00 and 583.14 eV represent Cr–O–N environments arising from oxygen incorporation, and the higher-energy features at 575.95 and 584.83 eV correspond to Cr–O states consistent with a surface oxide layer. Again, no metallic Cr contribution was observed, supporting the chemically stable incorporation of Cr into the nitride lattice.

The N 1s spectrum at Fig. S5c was deconvoluted into three components: the peak at 395.57 eV is assigned to Cr–N bonding, the dominant feature at 396.14 eV corresponds to Mn–N bonding, and the higher-binding-energy component at 397.50 eV is attributed to Mn–O–N/Cr–O–N configurations associated with mixed coordination arising from surface oxidation. Importantly, no higher-binding-energy peaks above 399 eV were observed, confirming the absence of molecular nitrogen species or nitrogen-related contaminants.

Taken together, these XPS results validate the formation of a chemically stable $Mn_3CrN$ phase with a rocksalt-type structure, in which nitrogen remains preferentially bonded to Mn and Cr in the lattice, and only minor surface oxidation, typical of nitride thin films, is present without compromising the bulk chemical integrity.

### E. Element-Specific X-ray Absorption Spectroscopy Measurements

X-ray absorption spectroscopy (XAS) was performed at both the N and O $K$-edges to examine the elemental composition and chemical integrity of the $Mn_3CrN$ thin film. Fig. S6a displays the N $K$-edge spectrum for the $Mn_3CrN$ film, which shows well-defined multiple absorption features between 400–440 eV, indicating the robust presence of nitrogen in the lattice. The prominent structure confirms the nitrogen-rich character of the film, supporting successful stoichiometric incorporation of nitrogen into the antiperovskite structure.

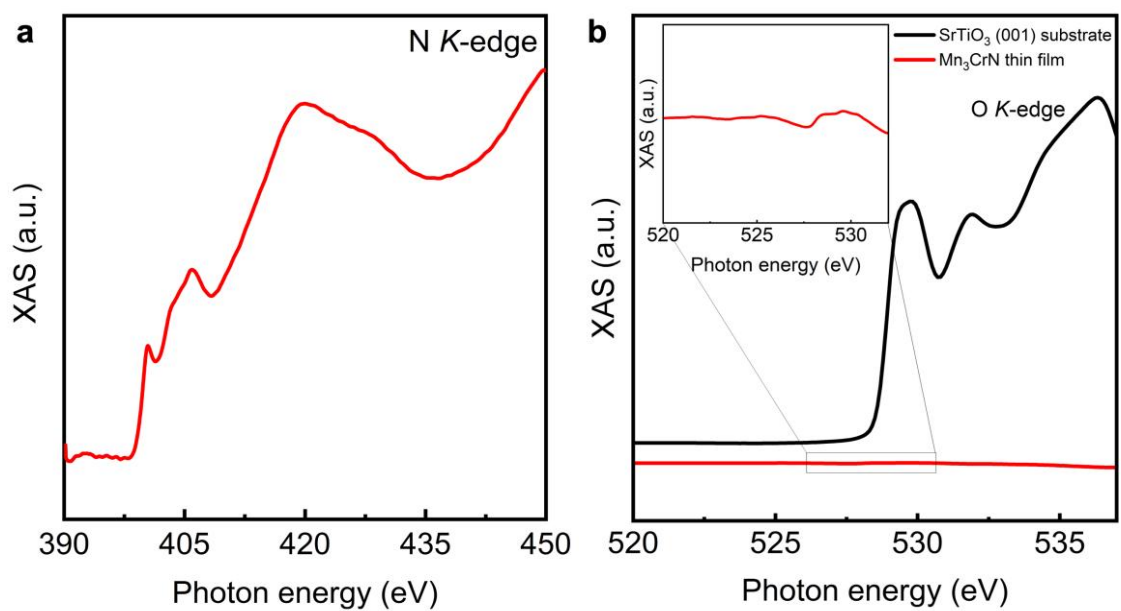

**Fig. S6 | X-ray absorption spectroscopy (XAS) of Mn₃CrN thin film. a**, N *K*-edge XAS spectrum of Mn₃CrN showing strong, well-defined multiplet features, confirming significant nitrogen incorporation consistent with the expected stoichiometry, **b**, O *K*-edge spectra comparing the Mn₃CrN film (red) and SrTiO₃ (001) substrate (black). The film shows negligible oxygen signal, suggesting only superficial oxygen presence, in stark contrast to the pronounced oxygen features from the substrate. Inset highlights the near-flat response of the film at the O *K*-edge.

In contrast, the O *K*-edge spectrum in Fig. S6b reveals a negligible response from the Mn₃CrN film in the 520–535 eV range. This absence of significant O-related features confirms minimal oxygen content in the film, suggesting only a superficial or adventitious oxygen layer, likely arising from brief air exposure. The strong O *K*-edge signal observed in the SrTiO₃ (001) substrate serves as a reference and clearly contrasts with the flat response from the film. These results validate the high purity and stoichiometric fidelity of the synthesized Mn₃CrN thin film with respect to their nitrogen and oxygen content.

### F. Temperature-dependent resistivity fitting of Mn₃CrN film

To examine the weak localization regime below ~35 K, the temperature-dependent resistivity data were fitted using the Mott variable range hopping (MVRH) model (see Fig. S7), expressed as

$$\rho = \rho_T \exp\left(\frac{T_M}{T}\right)^{1/3}$$
(S5)

where $\rho_T$ is a temperature-independent constant, $T_M$ is the Mott temperature, which relates to the localization strength $L_M$ and Bohr radius $r_{Bohr}$ as

$$T_M = \frac{18}{L_M^3 N(E_F) K_B}$$
(S6)

$$\bar{L}_M = r_{Bohr}\left(\frac{T_M}{T}\right)^{1/3}$$

(S7)

The term N(E$_F$) is the density of states (DOS) at the Fermi level.

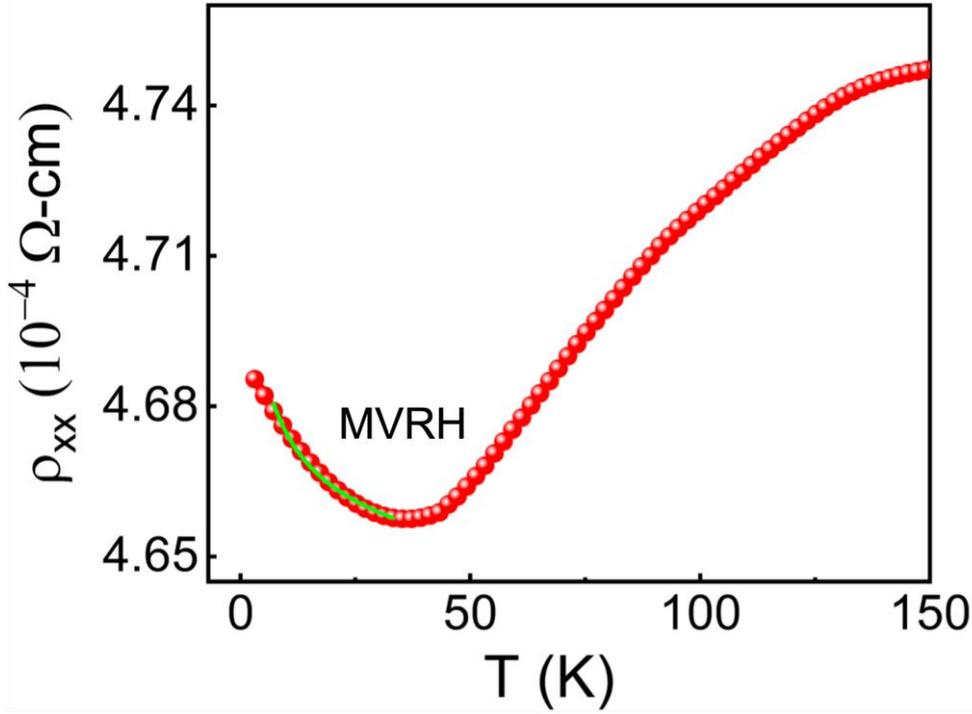

**Fig. S7** | The Mott variable range hopping (MVRH) fit of $\rho_{xx}$(T) below 35 K, confirming weak localization behaviour of Mn$_3$CrN film.

The MVRH fitting accurately reproduces the experimental $\rho_{xx}$(T) behaviour in the low-temperature regime, confirming that the conduction mechanism below ~35 K follows a 2D hopping process consistent with weak localization governed by quantum interference and disorder scattering. The fitting (Fig. S7) yields T$_M$ ≈ 1.35×10$^{-5}$ K, signifying a weak localization regime where electronic wavefunctions are only marginally localized.

**G. Intrinsic origin of the anomalous Hall effect revealed by scaling analysis**

The relationship between conductivity ($\sigma$) and resistivity ($\rho$) is given by the tensor inversion formula. For a 2D system or a film where the Hall resistivity is much smaller than the longitudinal resistivity ($\rho_{yx} \ll \rho_{xx}$), this leads to:

$$\rho_{yx}^{AH} \approx \sigma_{xy}^{AH} \cdot \rho_{xx}^2$$

(S8)

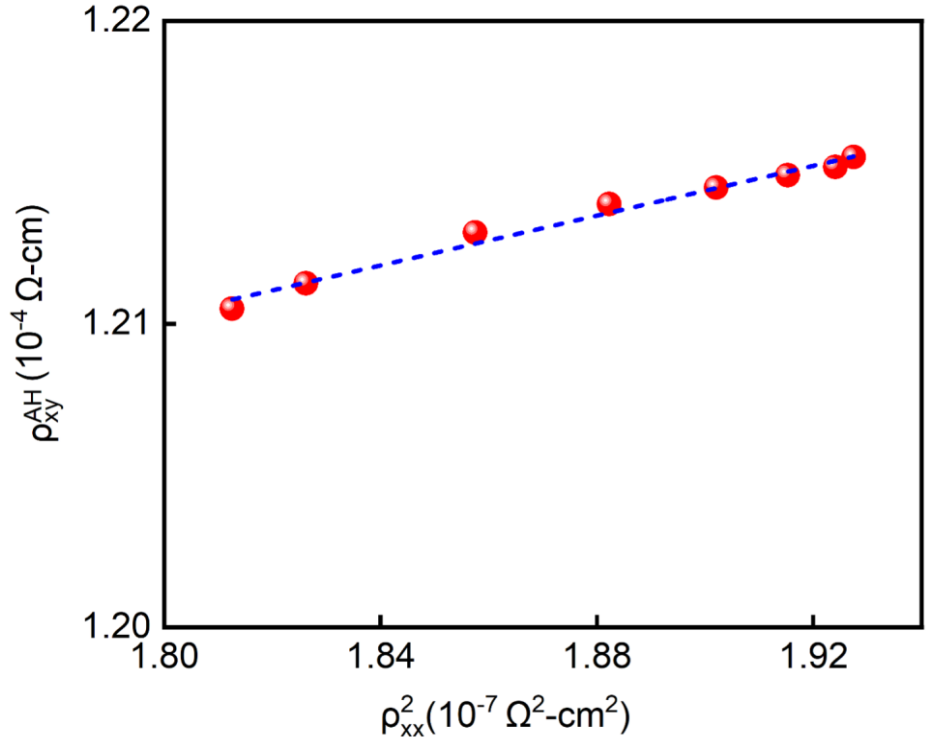

**Fig. S8** | Scaling relation between anomalous Hall resistivity ($\rho_{yx}^{AH}$) and longitudinal resistivity ($\rho_{xx}$) for Mn$_3$CrN.

Since the intrinsic anomalous Hall conductivity ($\sigma_{xy}^{AH,int}$) approximately constant with temperature (and therefore scattering rate), the temperature dependence of the anomalous Hall resistivity is governed by the temperature dependence of the longitudinal resistivity. By linear fitting, we obtained the $\sigma_{xy}^{AH,int}$ value around 41 Scm$^{-1}$ (see Fig. S8).

## H. Anomalous Hall Effect in Mn$_3$CrN Thin Films on STO Substrate

In addition to the AHC results for Mn$_3$CrN films deposited on (001) MgO substrates, AHC of Mn$_3$CrN films deposited on SrTiO$_3$ (STO) (001) substrates are also performed. Fig. S9a shows the field-dependent anomalous Hall conductivity ($\sigma_{xy}^{AH}$) measured at various temperatures, from 3 K to 375 K under out-of-plane magnetic field configuration. The hysteresis loop progressively narrows with increasing temperature, indicating a reduction in coercivity. Fig. S9b shows the remanent anomalous Hall conductivity ($\sigma_{xy}^{REM}$) as a function of temperature. A distinct peak is observed near 200 K, reminiscent of the temperature-induced vector spin chirality transition reported in MgO-based films in the main text. This again highlights a temperature-driven chirality transition in Mn$_3$CrN films deposited on STO substates.

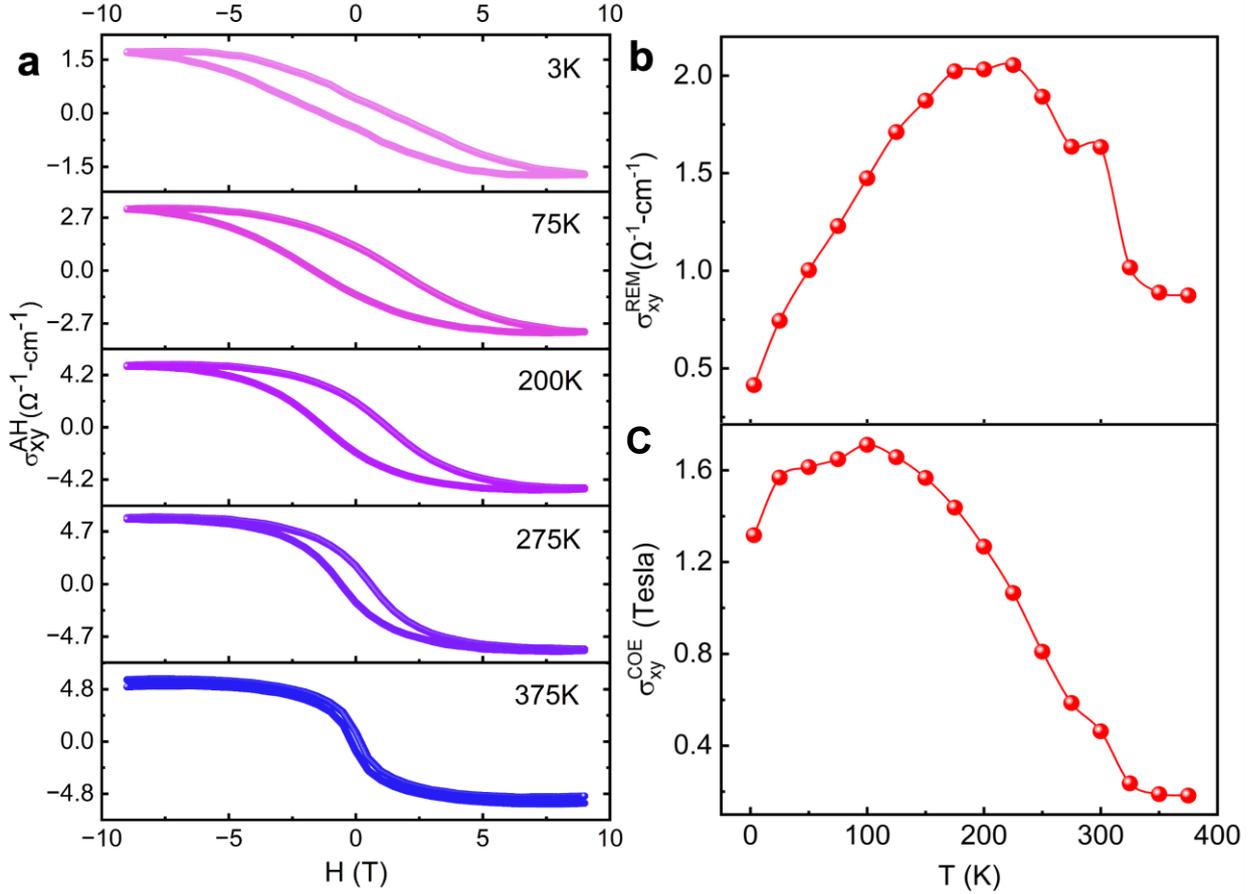

**Fig. S9 | Anomalous Hall Effect in Mn$_3$CrN Thin Films on STO Substrate. a**, Field-dependent anomalous Hall conductivity ($\sigma_{xy}^{AH}$) for Mn$_3$CrN/STO films measured under out-of-plane magnetic field at various temperatures, **b**, shows the temperature dependence of the remanent anomalous Hall conductivity ($\sigma_{xy}^{REM}$), showing a clear anomaly around 200 K, consistent with the spin chirality transition, **c**, shows the temperature dependence of the anomalous Hall coercivity ($\sigma_{xy}^{COER}$), indicating reduced coercive fields with temperature and a subtle kink near 75 K.

The temperature dependence of the coercivity of anomalous Hall conductivity ($\sigma_{xy}^{COE}$) in noncollinear antiferromagnetic Mn$_3$CrN, as depicted in Fig. S9c reveals a nonmonotonic profile. The coercivity initially increases at low temperature, peaks near 100 K, and then monotonically decreases with further temperature rise. This behaviour originates from the interplay between magnetic anisotropy energy and thermal fluctuations. At low temperatures, the magnetic anisotropy energy is maximized due to well-aligned spin configurations and strong spin-orbit/lattice coupling, resulting in enhanced domain wall pinning and coercivity.

As temperature increases, thermal agitation disrupts long-range magnetic order, leading to a reduction in the sublattice magnetization and consequently in the anisotropy energy. The anisotropy energy's temperature dependence can be quantitatively described by the phenomenological relation: [3]

$$E_{aniso}(T) = E_{aniso}(0) \left[1 - \left(\frac{T}{T_C}\right)^p\right] \tag{S9}$$

where $E_{aniso}(0)$ is the anisotropy energy at absolute zero, $T_C$ is the critical temperature (Curie or Néel), and $p$ is an exponent reflecting the material's characteristics. Above the coercivity peak, rising temperature increasingly facilitates domain wall motion and magnetization reversal, leading to the observed sharp decrease in coercivity. The combined effects of intrinsic anisotropy loss and enhanced thermal disorder thus govern the full temperature evolution of coercivity in Mn$_3$CrN.

These results confirm that the spin-chirality-mediated anomalous Hall response in Mn$_3$CrN is a robust feature of the system, observable on both STO and MgO substrates, with consistent magnetic and topological transitions.

I. Anomalous Hall Coercivity in Mn$_3$CrN/MgO

Fig. S10 presents the temperature-dependent anomalous Hall coercivity of Mn$_3$CrN films on MgO. The coercivity reaches a pronounced maximum near 100 K and decreases steadily with increasing temperature. The peak position closely matches that observed in STO-based films (see Fig. S9c).

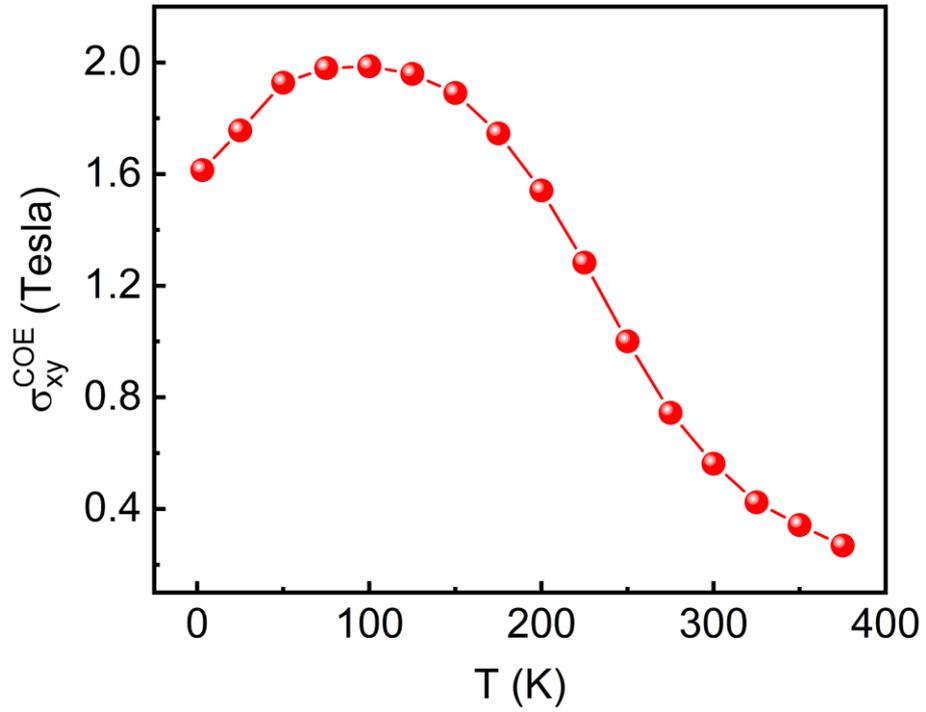

**Fig. S10** | Temperature dependence of anomalous Hall coercivity ($\sigma_{xy}^{COER}$) in Mn$_3$CrN/MgO films, showing a maximum near 100 K consistent with STO-based films (see Fig. S9c).

## J.  First principles DFT study of Mn₃CrN

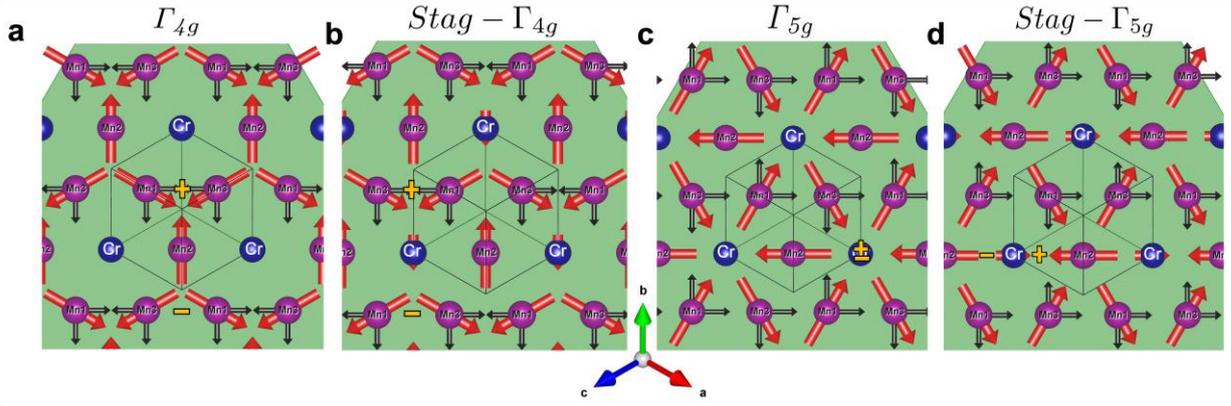

**Fig. S11 | Spin ordering and analysis of ∇·S in Mn₃CrN Kagome planes.** Spin textures for **a**, $\Gamma_{4g}$, **b**, staggered-$\Gamma_{4g}$ (s$\Gamma_{4g}$), **c**, $\Gamma_{5g}$, and **d**, staggered-$\Gamma_{5g}$ (s$\Gamma_{5g}$) phases are shown on the (111) Kagome plane. Mn spins (red arrows) are decomposed into longitudinal and transverse components, and points of source (+) and sink (−) in ∇·S are indicated (yellow symbols). Cr atoms (blue) occupy the hexagon centres. In the staggered configurations (**b,d**), the divergence paths intersect the Cr sites, introducing Cr moments. In contrast, for the non-staggered $\Gamma_{4g}$ and $\Gamma_{5g}$ phases (**a,c**), the divergence pattern avoids Cr positions, suppressing local moments at Cr-sites.

The divergence analysis (∇·S) provides a microscopic picture of how vector spin chirality couples the Mn and Cr sublattices (see Fig. S11). In the staggered phase ($\kappa = -1$), the non-uniform arrangement of spin sources (inward-pointing) and sinks (outward-pointing) within the (111) plane generates divergence lines that pass directly through the Cr sites. We describe this configuration as "monopole-like" because the source and sink behave as two distinct poles, with the lines of force between them mimicking those of magnetic monopoles. The Cr atoms, positioned along these lines, acquire small but finite moments that add extra stability to the staggered ground-state configuration. In contrast, in the non-staggered phase ($\kappa = +1$), the arrangement of sources and sinks is such that the divergence lines bypass the Cr sites, preventing the development of any moment. Thus, the presence or absence of Cr moments in Mn₃CrN is dictated by the chirality-dependent topology of the Mn spin texture. Experimentally, XMCD at 3.5 K detects no such Cr moment, since thermal deviations from the ground-state limit disrupt the monopole-like structure, while at higher temperatures the transition to the non-staggered phase removes the effect entirely.

### K. Energetic penalty of out-of-plane spin canting in Mn₃CrN

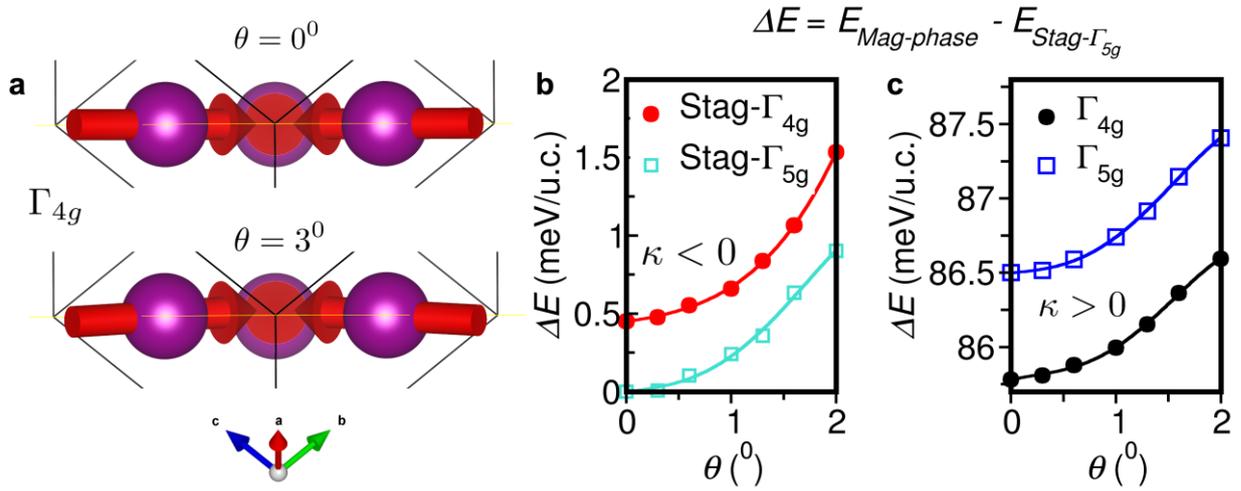

**Fig. S12 | Energies of out-of-plane spin canting in Mn₃CrN. a**, $\Gamma_{4g}$ spins in the coplanar ($\theta_{out} = 0°$) and canted ($\theta_{out} = 3°$) states. **b, c**, Energy difference ($\Delta E$) between the competing phases as a function of $\theta_{out}$, showing that canting increases the total energy for both $\kappa < 0$ and $\kappa > 0$ chirality states. The coplanar spin configuration is therefore the magnetic ground state.

To assess the robustness of the magnetic ground state, the energetic stability of out-of-plane spin canting is evaluated by introducing a finite polar angle $\theta_{out}$ with respect to the Kagome (111) plane. Fig. S12a illustrates the $\Gamma_{4g}$ spin configuration in the coplanar limit ($\theta_{out} = 0°$) compared with a 3° canted arrangement. The energy evolution (see Figs. S12b and S12c) reveals that for both chirality's, $\kappa = -1$ (staggered-$\Gamma$ phases) and $\kappa = 1$ (non-staggered-$\Gamma$ phases), the total energy rises monotonically with $\theta_{out}$. This systematic energy penalty indicates that canting away from the Kagome plane is not favorable, and the system strictly prefers the coplanar state.

Importantly, this result rules out the possibility of weak ferromagnetism originating from canting-induced uncompensated moments, as even small deviations from coplanarity carry a significant energetic cost. Thus, the coplanar spin texture within the (111) Kagome plane represents the true magnetic ground state of Mn₃CrN, while non-planar configurations are energetically unstable.

## L. Spin Rotation Continuum Between $\Gamma_{4g}$ and $\Gamma_{5g}$

To explore the relationship between $\Gamma_{4g}$ and $\Gamma_{5g}$ states, a systematic rotation of the Mn spins within the Kagome plane is performed, gradually evolving the configuration from $\Gamma_{4g}$ ($\theta = 0°$) to $\Gamma_{5g}$ ($\theta = 90°$), as illustrated in Fig. S13a. At each intermediate angle $\theta$, the total energy difference $\Delta E$ was calculated (see Fig. S13b).

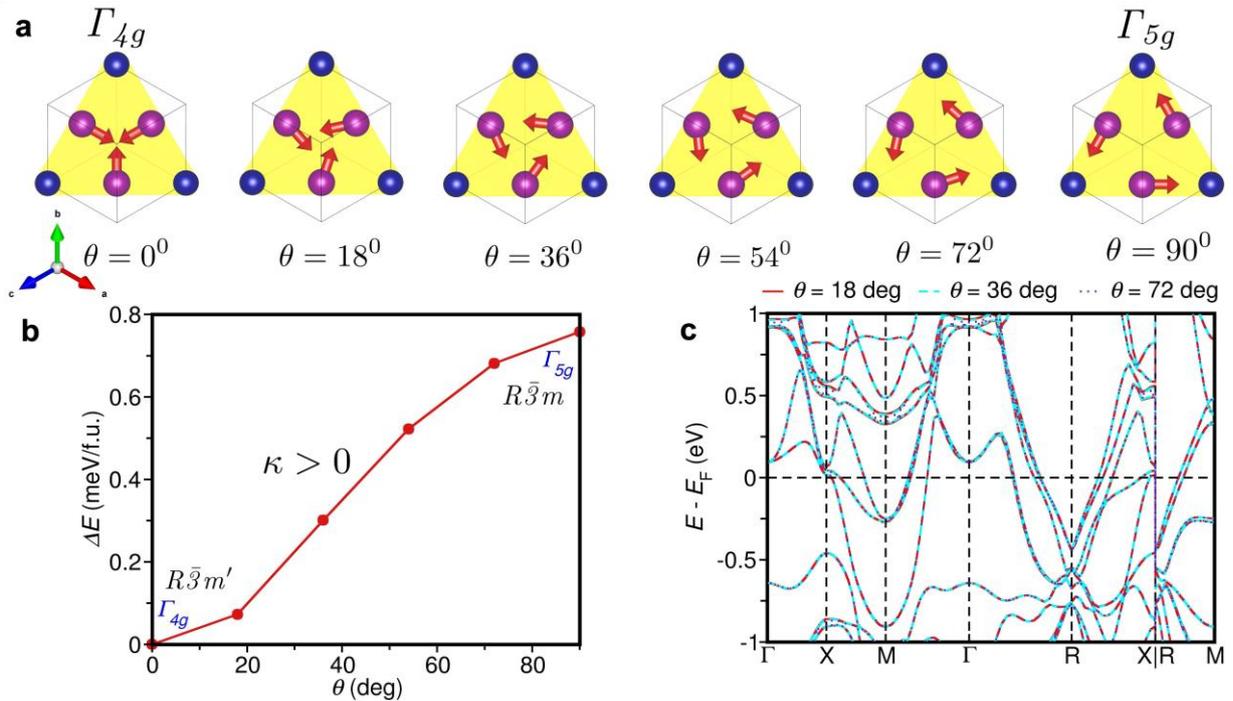

**Fig. S13 | Continuous rotation of spins connecting $\Gamma_{4g}$ and $\Gamma_{5g}$ configurations of Mn$_3$CrN. a**, Systematic evolution from $\Gamma_{4g}$ to $\Gamma_{5g}$ through consecutive spin rotations within the Kagome plane. **b**, Total energy variation ($\Delta E$) as a function of rotation angle $\theta$, showing only minor changes within the DFT error bar. **c**, Electronic band structures at representative angles ($\theta = 18°, 36°, 72°$), confirming full degeneracy of bands of all the intermediate configurations.

The variations in $\Delta E$ remain extremely small, within the error bar of the calculations, indicating that no meaningful energetic barrier exists along the rotation path. Consistent with this degeneracy, the calculated band dispersions (see Fig. S13c) show complete overlap for all intermediate configurations, with no detectable splitting or reconstruction. These results demonstrate that $\Gamma_{4g}$ and $\Gamma_{5g}$ are continuously connected through in-plane spin rotations without electronic or energetic distinction, reinforcing that the decisive factor differentiating them is the vector spin chirality rather than the underlying band structure.

**M. Spin Rotation Continuum Between staggered-$\Gamma_{4g}$ and staggered-$\Gamma_{5g}$**

The connection between staggered-$\Gamma_{5g}$ and staggered-$\Gamma_{4g}$ states are investigated further by gradually rotating the spins within the Kagome plane, as illustrated in Fig. S14a. The total energy variation with rotation angle θ (see Fig. S14b) remains below 0.5 meV/f.u., and indicates no meaningful energetic barrier along the rotation path.

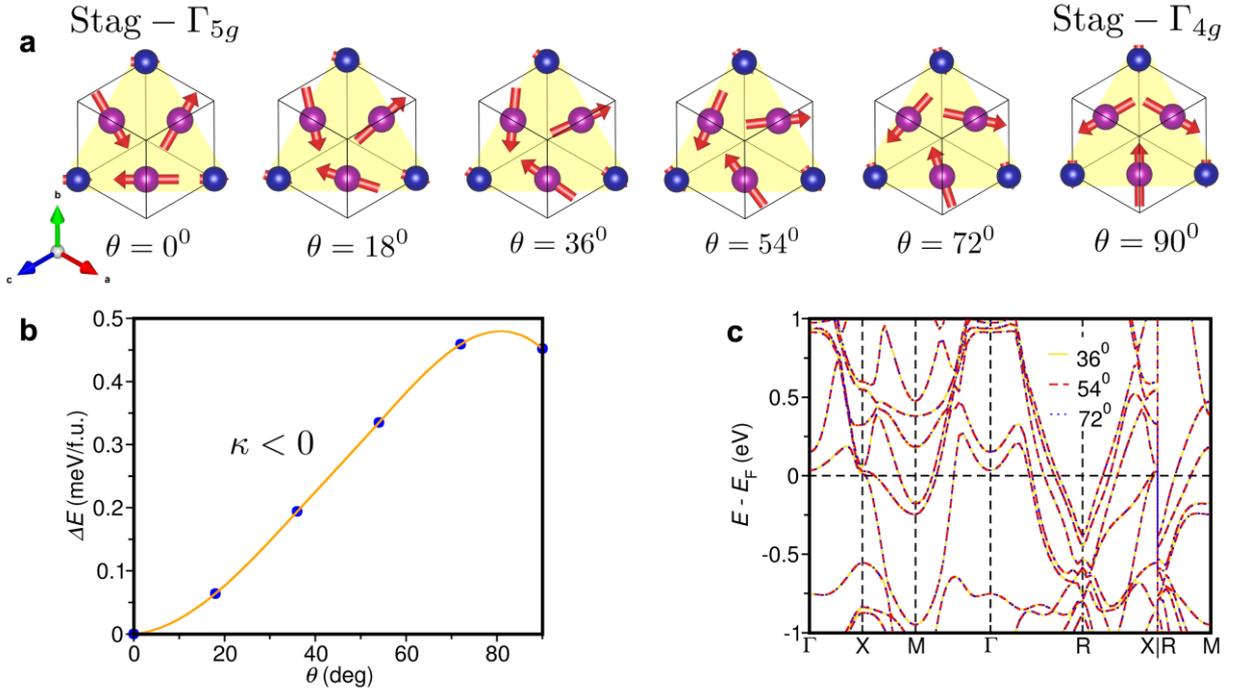

**Fig. S14 | Rotation path between staggered-$\Gamma_{5g}$ and staggered-$\Gamma_{4g}$. a**, Consecutive spin rotations evolving from staggered-$\Gamma_{5g}$ ($\theta = 0°$) to staggered-$\Gamma_{4g}$ ($\theta = 90°$). **b**, Total energy variation with rotation angle, showing only small changes within the sub-meV scale. **c**, Band structures at representative angles ($\theta = 36°$, $54°$, $72°$) exhibit full degeneracy, confirming electronic indistinguishability along the staggered rotation path.

Consistently, the electronic band dispersions at representative angles ($\theta = 36°$, $54°$, $72°$) show complete overlap (see Fig. S14c), demonstrating that the staggered phases are also electronically indistinguishable. These results confirm that the distinction between staggered-$\Gamma_{4g}$ and staggered-$\Gamma_{5g}$ states arises from spin chirality rather than electronic structure.

**Supplementary Information References**

1. Takenaka, K. *et al.* Magnetovolume effects in manganese nitrides with antiperovskite structure. *Sci. Technol. Adv. Mater.* **15,** 015009 (2014).
2. Katsuhiko Inaba. Introduction to XRD analysis of modern functional thin film using a 2-dimensional detector-(2) Analysis of epitaxial films. *Rigaku Journal* **33**, 10-14 (2017).
3. Phys, J. A. Temperature dependence of magnetic anisotropy constant in iron chalcogenide $Fe_3Se_4$: Excellent agreement with theories. *J. Appl. Phys.* **112**, 103905 (2012).